\documentclass[10pt]{article}
\usepackage{amsmath,amssymb,cite,epsfig}
\def \beq{\begin{equation}}
\def \eeq{\end{equation}}
\def \beqa{\begin{eqnarray}}
\def \eeqa{\end{eqnarray}}
\def \l{\left(}
\def \r{\right)}

\newcommand{\nn}{\nonumber}

\usepackage{xspace}

\begin{document}
\title{\begin{flushright}
\end{flushright}
{\bf Fluctuations and Correlations of Conserved Charges in the $(2+1)$ Polyakov Quark Meson Model}}
\author{Sandeep Chatterjee \footnote{sandeep@cts.iisc.ernet.in} \,and Kirtimaan A. Mohan\footnote{kirtimaan@cts.iisc.ernet.in}\\ 
\begin{small}
Centre for High Energy Physics, Indian Institute of Science, Bangalore, 560012, India.
\end{small}}
\date{\empty}
\maketitle

\begin{abstract}
We consider the $(2+1)$ flavor Polyakov Quark Meson Model and study the fluctuations (correlations) of conserved charges upto sixth (fourth) order. Comparison is made with lattice data wherever available and overall good qualitative agreement is found, more so for the case of the normalised susceptibilities.           The model predictions for the ratio of susceptibilities go to that of an ideal gas of hadrons as in Hadron Resonance Gas Model at low temperatures while at 
high temperature the values are close to that of an ideal gas of massless quarks. Our study provides a strong basis for the use of PQM as an
effective model to understand the topology of the QCD phase diagram.\\\\
PACS numbers: 05.40.-a, 12.38Aw, 12.38Mh, 12.39.-x
\end{abstract}
\maketitle
\section{Introduction}
The QCD medium is known to undergo phase transitions~\cite{QGP-Mc,QGP-Sv}. 
The QCD Lagrangian for $N_f$ flavors of massless quarks has a  
$SU_{L}\l N_f\r\times SU_{R}\l N_f\r$ global symmetry which spontaneously breaks 
into $SU_{V}\l N_f\r$ in the low energy hadronic vacuum by the formation of chiral condensate 
together with $N_f^2-1$ massless Goldstone bosons. In the opposite limit of infinite quark mass, 
QCD becomes a pure $SU\l3\r$ gauge theory. The low energy vacuum that possesses a center symmetry  
$Z\l3\r$ under the color gauge group (confined phase) gets spontaneously broken in the high 
temperature/baryon density regime (deconfined phase). The study of the various phases of strongly 
interacting matter has been the subject of intense theoretical 
research for some time now. With data coming in from RHIC and LHC, experiments too are not far 
behind and will provide us a unique opportunity to improve our understanding of the thermodynamics 
of strongly interacting matter.

At these temperatures and densities QCD is a strongly coupled theory and hence perturbative
methods to study the details of this phase transition fail. First principle
Lattice QCD (LQCD) Monte Carlo simulations give us important insights into various aspects of the
phase transition. However LQCD suffers from the sign problem at non zero baryon
density~\cite{LQCD-sign1,LQCD-sign2} and although several methods have been developed~\cite{LQCD-sign1,LQCD-sign2} 
to bypass the sign problem at small baryon chemical potential, a satisfactory solution
for all values still eludes us. Another approach is to study various 
phenomenological models whose phase diagrams possess the essential features of QCD. These models 
serve to complement LQCD computations and also provide us with an intuitive and physical understanding about
the behavior of the phases of strongly coupled matter in regions that are both accessible and inaccessible 
to LQCD.

QCD has three conserved charges: baryon number $B$, strangeness $S$ and electric charge $Q$. Thus 
the thermodynamic state of the bulk medium can be specified by four quantities: temperature $T$ and the 
chemical potentials corresponding to the conserved charges namely $\mu_B$, $\mu_Q$ and $\mu_S$. 
Effective QCD-like models show first order hadron-QGP phase transition for large $\mu_B$. This line 
of first order phase transition is expected to end at a critical end point (CEP) at finite $\mu_B$ 
since general symmetry arguments~\cite{symm} as well as 
LQCD~\cite{LQCD-crossover1,LQCD-crossover2,LQCD-crossover3,LQCD-crossover4} indicate a smooth 
crossover at $\mu_B=0$.
Divergent fluctuations and correlations are characteristics of critical behavior. In QCD, 
fluctuations and correlations of the conserved charges are expected to provide telltale signatures 
of the CEP~\cite{Stephanov:1998dy} and bring forth the structure of the QCD phase diagram. These 
fluctuations can also be extracted experimentally through event by event analysis making them 
important observables in understanding the nature of strongly interacting matter 
\cite{Stephanov:event,C_BS}.

There have been numerous studies of the QCD correlators, both in lattice as well as 
models. LQCD with 2 flavors found 
the fluctuations to rise steeply~\cite{LQCD-steep1,LQCD-steep2} while 
the higher order susceptibilities exhibited non-monotonic behaviour~\cite{LQCD-nonm1,LQCD-nonm2} 
in the crossover region.  Similar studies have been performed in the 2 flavor Polyakov Nambu Jona Lasinio (PNJL) model~\cite{PNJLsus2f1,PNJLsus2f2,PNJLsus2f3,PNJLsus2f4,PNJLsus2f5}, 2 flavor Quark Meson (QM) model~\cite{QMsus2f},
2 flavor Polyakov Quark Meson (PQM) model~\cite{PQMsus2f} as well as in Hard Thermal Loop approximation~\cite{HTLsus1,HTLsus2,HTLsus3}. 
Susceptibilities were also computed in an improved version of the 2 flavor PQM model with the addition of the frequently 
neglected fermion vacuum fluctuations~\cite{PQMsus2f_RG1,PQMsus2f_RG2}. For $\l2+1\r$ flavors,
there are LQCD measurements~\cite{LQCD-steep3,LQCD-steep4,HotQCDlat2,WBlat} as well as model studies in 
PNJL~\cite{PNJLsus2p1f1,PNJLsus_china1,PNJLsus_china2,PNJLsus_kol1,PNJLsus_kol2} 
and PQM~\cite{PQMsus2p1f1,PQMsus2p1f2}. Recently, higher order generalised susceptibilities
and their sign structure in the phase diagram was studied in the $\l2+1\r$ PQM model~\cite{sus_sign}.

In this paper, we study the diagonal and off-diagonal susceptibilities of 
the conserved charges  $B$, $Q$ and $S$ in the $(2+1)$ PQM model at zero chemical 
potentials with the inclusion of the contribution from the fermionic vacuum fluctuations.
The significance of the vacuum term was recently pointed out in the case of 2 
flavours~\cite{PQMsus2f_RG1} where it was shown that the order of the phase transition 
in the massless limit changes from first order to crossover on adding the vacuum term. 
In our earlier study~\cite{PQMVT-2+1}, we have shown the significance of this vacuum term 
on the thermodynamics in the case of $(2+1)$ flavors. The phase transition region became 
smoother which led to better agreement with lattice. Here we report our investigation on the 
susceptibilities of this model and how they compare with results from the HotQCD~\cite{HotQCDlat2} 
and Wuppertal-Budapest (WB)~\cite{WBlat} LQCD groups. Our paper is organised as follows: In Sec. 
\ref{formalism}, we discuss the details of the PQM model and its parameters. 
In Sec. \ref{result}, we present our results and compare with those obtained in LQCD. 
Finally, in Sec. \ref{conclusion} we summarise and conclude.

\section{FORMALISM}
\label{formalism}
\subsection{Model}
\label{model}
We had formulated the $(2+1)$ flavor PQM with the inclusion of the vaccum term in~\cite{PQMVT-2+1}. 
Here we work with the same model. The relevant thermodynamic potential at a temperature 
$T$ and chemical potentials $\mu_B$, $\mu_Q$ and $\mu_S$ in the mean field approximation can be written as~\cite{PQMVT-2+1}
\beq
 \Omega \l T, \mu_B, \mu_Q, \mu_S \r = {\cal U_{\text{M}}}\l \sigma_u, \sigma_d, \sigma_s \r +
 {\cal U_{\text{Poly-VM}}} \l\Phi, \bar{\Phi}, T \r+ \Omega_{\bar{q}q}\l T,\mu_B, \mu_Q, \mu_S, \sigma_u, \sigma_d, \sigma_s, \Phi, \bar{\Phi} \r
\label{eq.omega}
\eeq
Here $\sigma_f$ denotes a chiral condensate of the quark with flavor $f$ (up(u), down(d) and strange(s)). 
${\cal U_{\text{M}}}\l \sigma_u, \sigma_d, \sigma_s \r$
is the contribution from the mesonic potential given by
\beqa
\label{eq.mesop}
 {\cal U_{\text {M}}}(\sigma_{u},\sigma_{d},\sigma_{s})&=&\frac{m^{2}}{2}\left( \frac{1}{2}\left(\sigma_{u}^{2} +\sigma_{d}^{2}\right)
 + \sigma_{s}^{2}\right) -\frac{h_{u}}{2} \sigma_{u} -\frac{h_{d}}{2} \sigma_{d} -h_{s} \sigma_{s}
 - \frac{c}{2 \sqrt{2}} \sigma_{u} \sigma_{d} \sigma_{s}\nn\\
 && + \frac{\lambda_{1}}{4} \left(\sigma_{u}^{2}+\sigma_{d}^{2}\right) \sigma_{s}^{2}+
\frac{\lambda_{1} + \lambda_{2}}{16} \left(\sigma_{u}^{4}+\sigma_{d}^{4} + 4 \sigma_{s}^{4}\right) +
\frac{\lambda_{1}}{8}\sigma_{u}^{2}\sigma_{d}^{2}
\eeqa

The Polyakov loop potential ${\cal U_{\text{Poly-VM}}} \l \Phi , \bar{\Phi} \r$ with the Jacobian Van der Monde term 
is given by~\cite{VM}
\beq
 \frac{{\cal U_{\text{Poly-VM}}}\left(\Phi,\bar{\Phi}, T \right)}{T^4} =
 \frac{{\cal U_{\text{Poly}}}\left(\Phi,\bar{\Phi}, T \right)}{T^4} -
  \kappa \log\left[1-6\Phi\bar{\Phi}+4\l\Phi^3+{\bar{\Phi}}^3\r-3\l\Phi\bar{\Phi}\r^2\right].
\label{eq.VMpot}
\eeq
Here ${\cal U_{\text{Poly}}}$ is the Ginzburg-Landau type potential given by~\cite{ratti}
\beq
\frac{{\cal U_{\text{Poly}}}\left(\Phi,\bar{\Phi}, T \right)}{T^4}=-\frac{b_2(T)}{2}\Phi\bar{\Phi}-
  \frac{b_3}{6}\l\Phi^3+{\bar{\Phi}}^3\r+\frac{b_4}{4}\l\Phi\bar{\Phi}\r^2
\label{eq.polpot}
\eeq
with
\beq
b_2\l T\r=a_0+a_1\frac{T_0}{T}+a_2\l\frac{T_0}{T}\r^2+a_3\l\frac{T_0}{T}\r^3
\label{b2}
\eeq
where
\begin{eqnarray}
&& a_0 = 6.75\ , \qquad a_1= -1.95\ , \qquad b_3= 0.75 \nn \\ 
&& a_2 =2.625\ ,  \qquad a_3=-7.44\ ,   \qquad b_4=7.5 \nn 
\end{eqnarray}

The quark/antiquark contribution is given by 
\beq
\Omega_{\bar{q}q}\l T,\mu_B, \mu_Q, \mu_S, \sigma_u, \sigma_d, \sigma_s, \Phi, \bar{\Phi} \r =\Omega^{\text{v}}_{\bar{q}q}\l \sigma_u, \sigma_d, \sigma_s \r+
 \Omega^{\text{th}}_{\bar{q}q}\l T,\mu_B, \mu_Q, \mu_S, \sigma_u, \sigma_d, \sigma_s, \Phi, \bar{\Phi} \r
\label{omegaqq}
\eeq
$\Omega^{\text{th}}_{\bar{q}q}$ is the quark/antiquark contribution due to thermal
fluctuations
\beqa
\Omega^{\text{th}}_{\bar{q}q}\l T,\mu_B, \mu_Q, \mu_S, \sigma_u, \sigma_d, \sigma_s, \Phi, \bar{\Phi} \r&=&-2T \sum_{f=u,d,s} \int \frac{d^3 p}{(2\pi)^3}
 \ln \Big[ 1 + 3\Phi e^{ -E_{f}^{+} /T} +3 \bar{\Phi}e^{-2 E_{f}^{+}/T} +e^{-3 E_{f}^{+} /T}\Big]\nn\\
&&-2T \sum_{f=u,d,s} \int \frac{d^3 p}{(2\pi)^3}
 \ln \Big[ 1 + 3\bar{\Phi} e^{ -E_{f}^{-} /T} +3 \Phi e^{-2 E_{f}^{-}/T} +e^{-3 E_{f}^{-} /T}\Big]
\label{eq.omegath}
\eeqa
where
\beq
E_{f}^{\pm} =E_f \mp \mu_f\l \mu_B, \mu_Q, \mu_S \r
\label{eq.ef}
\eeq
$\mu_f$ denote the quark chemical potentials $\mu_u$, $\mu_d$ and $\mu_s$. They are related 
to $\mu_B$, $\mu_Q$ and
$\mu_S$ by the following transformations
\beqa
\mu_u &=& \frac{\mu_B}{3} + \frac{2 \mu_Q}{3}\nn\\
\mu_d &=& \frac{\mu_B}{3} - \frac{\mu_Q}{3}\nn\\
\mu_s &=& \frac{\mu_B}{3} - \frac{\mu_Q}{3} - \mu_S\nn\\
\label{eq.muqtomuH}
\eeqa
In (\ref{eq.ef}), $E_f$ is the single particle energy of a quark/antiquark.
\beq
E_f = \sqrt{p^2 + m{_f}{^2}}
\eeq
and $m_f$ is the mass of a quark/antiquark with flavor $f$ given by
\beqa
m_f &=& \frac{g}{2}\sigma_f,\,\,f \in u,d\nn\\
 &=& \frac{g}{\sqrt{2}}\sigma_f,\,\,f \in s
\eeqa
$\Omega^{\text{v}}_{\bar{q}q}$ is the vacuum term
\beqa
\Omega^{\text{v}}_{\bar{q}q}\l \sigma_u, \sigma_d, \sigma_s \r&=&-2N_c\sum_{f=u,d,s}\int \frac{d^3 p}{(2\pi)^3}E_f \nn\\
&=&-\frac{N_c}{8 \pi^2}\sum_{f=u,d,s}m_f^4\log\left[\frac{m_f}{\Lambda}\right]
\label{omegavac}
\eeqa
where $\Lambda$ is the regularisation scale. In~\cite{PQMVT-2+1}, we have shown that at the mean
field level, $\Omega$ and hence all the thermodynamic quantities are independent of the choice
of $\Lambda$. The model parameters: $m^2$, $\lambda_1$, $\lambda_2$, $c$, $g$, $h_u$, $h_d$ and $h_s$
are determined by establishing the vacuum properties. The parameters of $b_2(T)$, $b_3$, $b_4$, $T_0$
and $\kappa$ are obtained by fitting to the lattice data~\cite{ratti,PQMVT-2+1}. Two sets of model parameters 
were obtained in our earlier work~\cite{PQMVT-2+1}, namely ModelHotQCD and ModelWB. Here we work with 
the same model parameters. In Table \ref{tb.param}, we list the values of all the parameters that have 
been used in this work.
\begin{table}[htb]
\begin{center}
\begin{tabular}{|c|c|c|c|c|c|c|c|c|c|}
\hline
Model&$m^2$ $[\text{MeV}^2]$&$\lambda_1$&$\lambda_2\l\Lambda\r$&$c$&$h_u=h_d$ 
$[\text{MeV}^3]$&$h_s$ $[\text{MeV}^3]$&$g$&$T_0$&$\kappa$\\
\hline
\hline
ModelHotQCD & 80647.587 & -8.165 & 138.45 & 4801.95 & $1.785\times10^6$ & $3.805\times10^7$ & 6.5 & 210 & 0.1\\
\hline
ModelWB & 80647.587 & -8.165 & 138.45 & 4801.95 & $1.785\times10^6$ & $3.805\times10^7$ & 6.5 & 270 & 0.2\\
\hline
\hline
\end{tabular}
\end{center}
\caption{The parameter sets obtained with $\Lambda=200$ MeV.}
\label{tb.param}
\end{table}

The mean field values of the condensates $\sigma_u$, $\sigma_d$ and $\sigma_s$ and the Polyakov loop
variables $\Phi$ and $\bar{\Phi}$ are determined by numerically solving the following set of equations
\beq
  \frac{ \partial \Omega}{\partial
      \sigma_u} =\frac{ \partial \Omega}{\partial
      \sigma_d} = \frac{ \partial \Omega}{\partial \sigma_s} 
      = \frac{ \partial \Omega}{\partial \Phi}
      = \frac{\partial \Omega}{\partial \bar{\Phi}}
   = 0
\label{eq.gap}
\eeq
\subsection{Susceptibilities}
\label{susceptibilities}
The pressure $P$ is given by
\beq
\label{eq.pressure}
P\l T, \mu_B, \mu_Q, \mu_S \r=-\Omega \l T, \mu_B, \mu_Q, \mu_S \r
\eeq
In order to find the generalised susceptibilities of the conserved charges, one has to 
take appropriate derivatives of $P$
\beq
\chi^{BQS}_{ijk}=\frac{\partial^{i+j+k}(P/T^4)}{\partial \l \mu_B/T\r^i\partial \l \mu_Q/T\r^j\partial 
\l \mu_S/T\r^k}
\label{eq.sus}
\eeq
In this work, we perform all our computations at zero chemical potentials. 
The baryon number, electric charge, and strangeness densities vanish at zero chemical potentials while the higher order derivatives with $\l i+j+k\r$ even are nonzero. In this work, we compute 
$\chi^{BQS}_{ijk}$ upto $\l i+j+k\r=6$ order. The derivatives in (\ref{eq.sus}) have been computed 
using the algorithmic differentiation techniques available in ADOL-C~\cite{ADOLC} which allow us to 
compute higher order derivatives efficiently and without additional truncation errors.
\section{Result}
\label{result}
In the low $T$ regime, the condensates have large values which impart large mass
to the relevant degrees of freedom. Thus all susceptibilities are small in the low $T$ regime. In the crossover regime, the condensates start to melt and thus the relevant degrees of freedom become lighter. Generically, this results in a higher value of the susceptibilities as one approaches $T_c$. These quantities have been extensively computed on the lattice~\cite{LQCD-nonm1,LQCD-nonm2,LQCD-steep1,LQCD-steep2,LQCD-steep3,LQCD-steep4,HotQCDlat2,WBlat}. In the high temperature limit when $T$ is the only relevant scale, one expects the system to behave like a Stefan-Boltzmann (SB) gas of 3 flavors of massless quarks ( we don't consider the gluons as they do not carry any B, Q and S quantum numbers and hence don't contribute to the susceptibilities in the high temperature limit ) whose pressure is given by
\beq
\frac{P^{SB}}{T^4}=\sum_{f=u,d,s}\left[\frac{7\pi^2}{60}+\frac{1}{2}\l\frac{\mu_f}{T}\r^2+
                   \frac{1}{4\pi^2}\l\frac{\mu_f}{T}\r^4\right]
\label{eq.pSB}
\eeq
\begin{table}
\begin{center}
\begin{tabular}{|p{0.5in}|p{0.5in}|p{0.5in}|p{0.5in}|}
\hline
\hline
 $X $& B & Q & S\\
\hline
$\chi_2^{X}$ & $1/3$ & $2/3$ & 1 \\
$\chi_4^{X}$ & $2/9\pi^2$ & $4/3\pi^2$ & $6/\pi^2$ \\
\hline
\hline
XY & BQ & BS & QS\\
\hline
$\chi_{11}^{XY}$ & 0 & $-1/3$ & $1/3$ \\
\hline
$\chi_{22}^{XY}$ & $4/9\pi^2$ & $2/3\pi^2$ & $2/3\pi^2$ \\
\hline
$\chi_{31}^{XY}$ & $0$ & $-2/9\pi^2$ & $2/9\pi^2$ \\
\hline
$\chi_{13}^{XY}$ & $4/9\pi^2$ & $-2/\pi^2$ & $2/\pi^2$ \\
\hline
\hline
XYZ & BQS & QSB & SBQ\\
\hline
$\chi_{211}^{XYZ}$ & $2/9\pi^2$ & $-2/9\pi^2$ & $-2/3\pi^2$ \\
\hline
\hline
\end{tabular}
\end{center}
\caption{The SB values of the susceptibilities and correlations.}
\label{tb.SB}
\end{table}
The corresponding SB limit values for quadratic and quartic fluctuations of conserved charges as
well as their correlations are given in Table \ref{tb.SB}. On the other hand, at the low $T$ limit, an ideal gas of hadrons as in hadron resonance gas models (HRGM) is expected to describe the thermodynamics and susceptibilities. Latest LQCD data from WB~\cite{WBlat} exhibit the above expectations quite clearly. Since the PQM model has quark degrees of freedom, in the the high $T$ limit, one expect that the susceptibilities
will go to that of an ideal SB gas of quarks. On the contrary, in the low $T$ side, since the PQM model does not have baryons (there are only 'three quark' states which mimic 
baryonic degrees of freedom in the low $T$ limit in the model~\cite{chi4Btochi2B}), one 
should not expect the model predictions of the susceptibilities to match with those of HRGM. 
However, in case of ratios of susceptibilities, we argue below that that in the low $T$ limit the PQM results match with that of HRGM~\cite{chi4Btochi2B}.

Let us first look at the susceptibility ratios for low temperatures in PQM. At low $T$, $\phi$ and $\bar{\phi}$ are nearly $0$ and hence from (\ref{eq.omegath}) we find that the excitations of one and 
two quark states are suppressed. Thus, we may approximately write the thermal quark-antiquark contribution to the pressure as~\cite{chi4Btochi2B}
\beq
\frac{P^{th}_{q\bar{q}}\l T,\mu\r}{T^4}\sim\frac{d_q}{27 T^2}\l \frac{3m_q}{T}\r^2K_2\l\frac{3m_q}{T}\r\text{cosh}\l\frac{3\mu_q}{T}\r
\label{eq.pressthqaq}
\eeq
which is same as that of a non-interacting gas of paticles and anti-particles with baryon 
number 1 and -1 respectively and mass $3m_q$. Here $d_q$ is the effective degeneracy factor which depends on the particular susceptibility one is looking at: $d_q$ is 2 when the contribution from the non strange quark sector is to be considered while $d_q$ is 1 where the strange quark sector is important. Thus we find in the low $T$ limit there is a nice factorization and one could write (\ref{eq.pressthqaq})
\beq
\frac{P^{th}_{q\bar{q}}\l T,\mu\r}{T^4}\sim f\l\frac{m_q}{T}\r\text{cosh}\l\frac{3\mu_q}{T}\r
\label{eq.prPQM}
\eeq
Now we shall see that in the low $T$ limit, a similar factorization also occurs in the HRGM framework whose pressure is given by that of an ideal gas of hadrons
\beq
 p^{\text{HRG}}/T^4=\frac{1}{VT^3}\sum_{i\in\text{hadrons}}\ln Z_i\l T,V,\mu_B,\mu_Q,\mu_S \r
\label{pHRG}
\eeq
where
\beq
\ln Z_i = \frac{aVd_i}{2\pi^2}\int_0^{\infty}dpp^2\ln\l1 + az_ie^{-\varepsilon_i/T} \r
\label{zi}
\eeq
is the contribution from the $i$th hadron with $a=\mp 1$ depending on whether it is meson or baryon respectively, $\varepsilon_i=\sqrt{p^2+m_i^2}$ where $m_i$ is the mass of the $i$th hadron with degeneracy
factor $d_i$ and fugacity
\beq
z_i=\exp \l\l B_i\mu_B+Q_i\mu_Q+S_i\mu_S\r/T\r
\eeq
The sum in (\ref{pHRG}) is usually over all hadrons from the particle data book with
masses $m_i \leq 2.5$ GeV. However, recently the significance of the yet undiscovered heavier hadrons on the heavy ion phenomenology and freezeout conditions were pointed out in~\cite{hag1,hag2}. Expanding the fugacity term one can write (\ref{zi}) in the following series form
\beq
\ln Z_i=\frac{VT^3}{\pi^2}d_i\l\frac{m_i}{T}\r^2\sum_{l=1}^{\infty}\l-a\r^{l+1}l^{-2}K_2\l lm_i/T\r \text{cosh}\l l\l B_i\mu_B+Q_i\mu_Q+S_i\mu_S\r/T\r
\label{lnzi}
\eeq
Thus (\ref{pHRG}) becomes
\beq
 p^{\text{HRG}}/T^4=\sum_i \frac{d_i}{\pi^2}\l\frac{m_i}{T}\r^2\sum_{l=1}^{\infty}\l-a\r^{l+1}l^{-2}K_2\l lm_i/T\r 
\text{cosh}\l l\l B_i\mu_B+Q_i\mu_Q+S_i\mu_S\r/T\r
\label{eq.phrg}
\eeq
Note that since $-a=1$ for a meson, all the terms
 of the series in~(\ref{eq.phrg}) are positive while in the case of a baryon, it is an 
 alternate series.
For sufficiently small temperature compared to the mass of the hadron one can approximate the series in $l$ by its leading
term which is the Boltzmann approximation. Also the dominant contribution will come from the lightest relevant hadron $h_l$,
\beq
p^{\text{HRG}}/T^4 \approx \frac{d_l}{\pi^2}\l\frac{m_l}{T}\r^2K_2\l\frac{m_l}{T}\r\text{cosh}\l \l B_l\mu_B+Q_l\mu_Q+S_l\mu_S\r/T\r
\label{eq.phrgMB}
\eeq
Thus, once again we find a factorization in the low $T$ limit in HRGM similar to 
(\ref{eq.prPQM}) for the case of PQM. This factorization leads to the fact that in the 
low $T$ limit ratios of susceptibilities are independent of the details of the mass spectrum both in the case of HRGM as well as PQM. Thus as long as the relevant degrees of freedom in both PQM and HRGM have the same quantum numbers, their ratios of susceptibilities should match.

It is easy to check from (\ref{eq.phrgMB}) that ratios like ${\chi}^X_i/{\chi}^X_j$ become simply $X_l^{i-j}$ where $X \in B, Q, S$ and $X_l$ are the $B, Q, S$ quantum numbers of $h_l$. Since the lightest hadrons in any sector are singly charged, $X_l^{i-j}$ is just unity and independent of the details of the hadron spectrum. However, in case of $S$ and $Q$ sectors, the contribution from multiply strange or charged hadrons enhances the higher order fluctuations compared to that of the second order and leads to deviation from unity for slightly higher temperatures. The light mass of the pions result in the breakdown of the Boltzmann approximation at slightly higher temperatures which in turn affects the ratios in the Q sector. As seen from (\ref{eq.phrg}), in the case of pion which is a meson, the quantum corrections are all positive and higher the order of the susceptibility larger is the correction. This enhances ratios likes $\chi^Q_4/\chi^Q_2$ and $\chi^Q_6/\chi^Q_2$.

We will now present the generalised susceptibilities and correlations computed in PQM at zero chemical potentials. We will also
present the LQCD data from HotQCD and WB groups for comparison. We find overall good qualitative agreement between model predictions
and LQCD. Particularly in the case of ratios, the PQM predictions interpolate between the low temperature HRGM values and the high temperature SB values.

\subsection{$\l i+j+k \r = 2$}
There are six second order susceptibilities: three diagonal susceptibilties 
$\chi_2^B$, $\chi_2^Q$, $\chi_2^S$ and three off-diagonal ones, $\chi_{11}^{BQ}$, $\chi_{11}^{QS}$ 
and $\chi_{11}^{BS}$.
\begin{figure}
  \scalebox{0.75}{\includegraphics{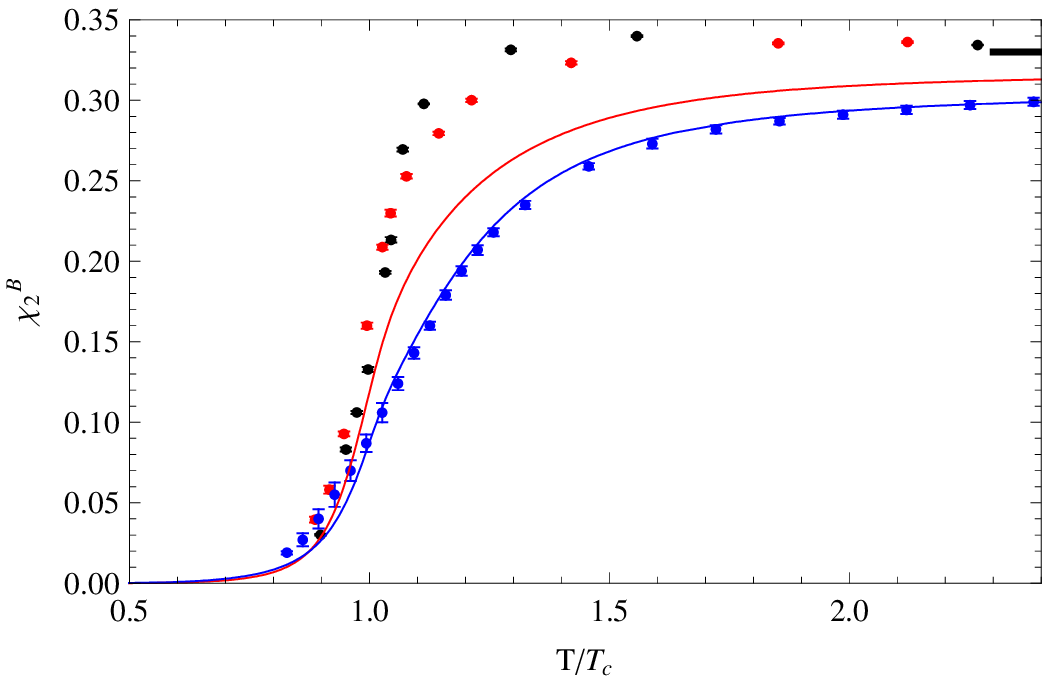}}
  \scalebox{0.75}{\includegraphics{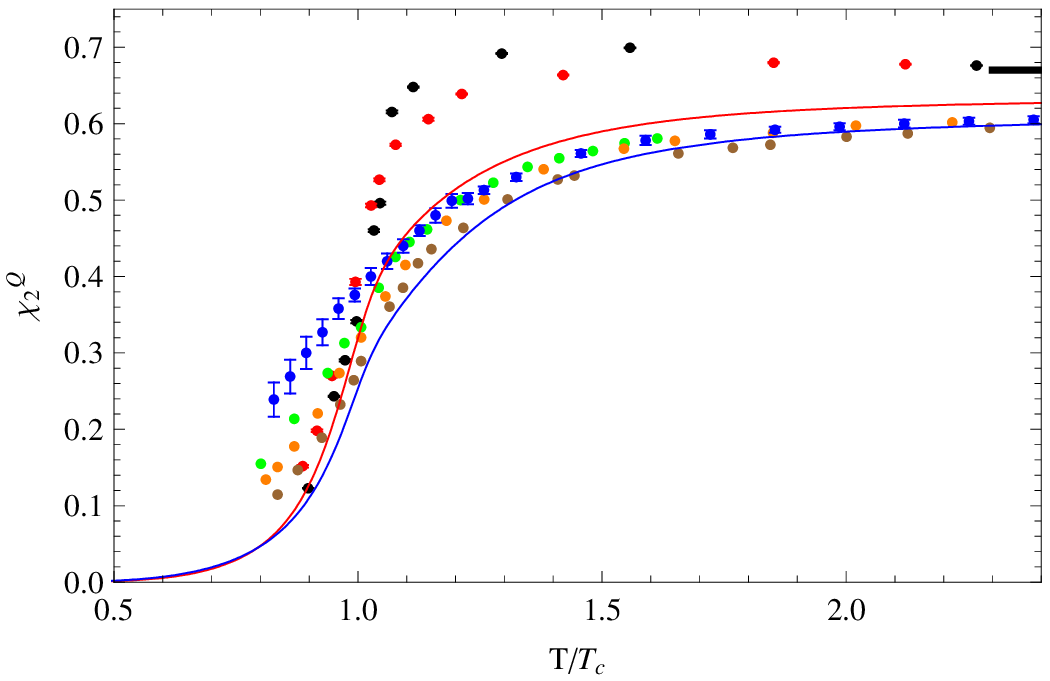}}\\
  \scalebox{0.75}{\includegraphics{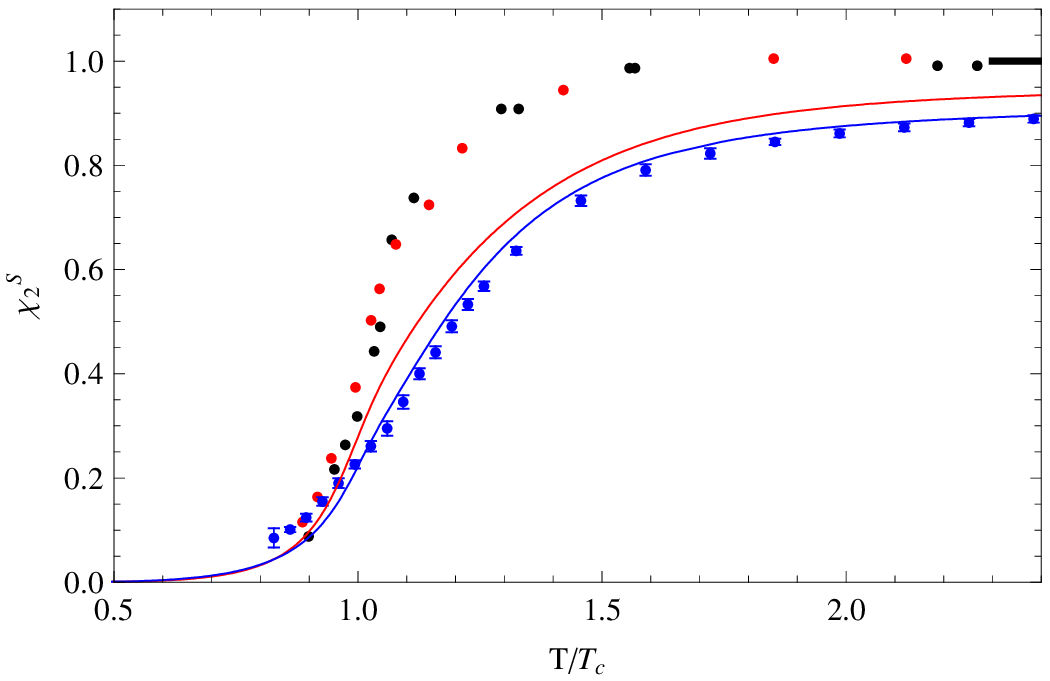}}
  \scalebox{0.75}{\includegraphics{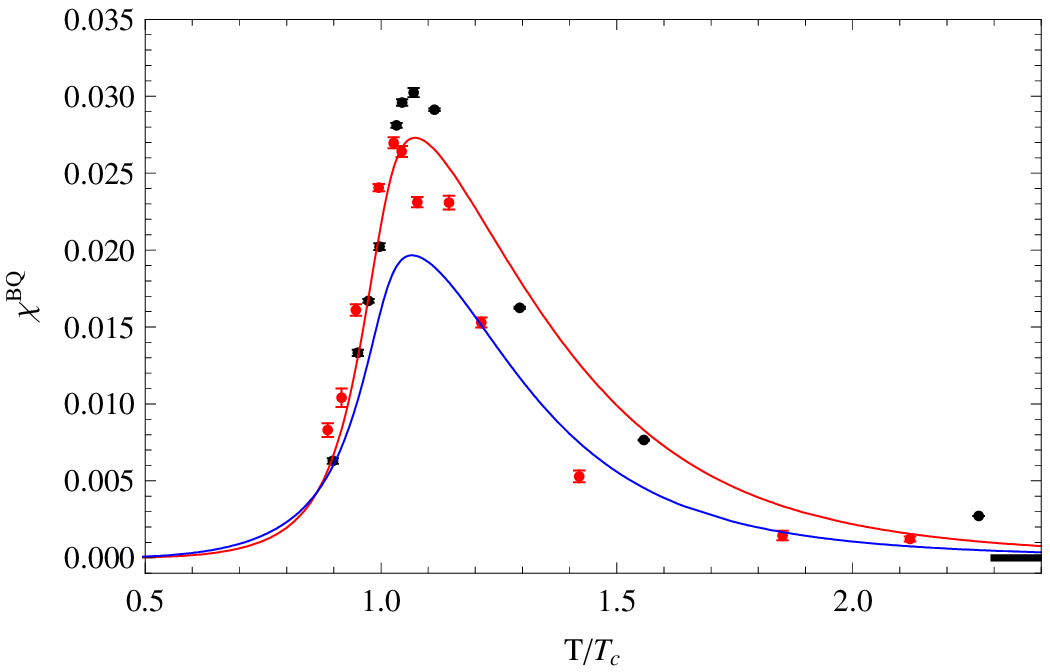}}\\
  \scalebox{0.75}{\includegraphics{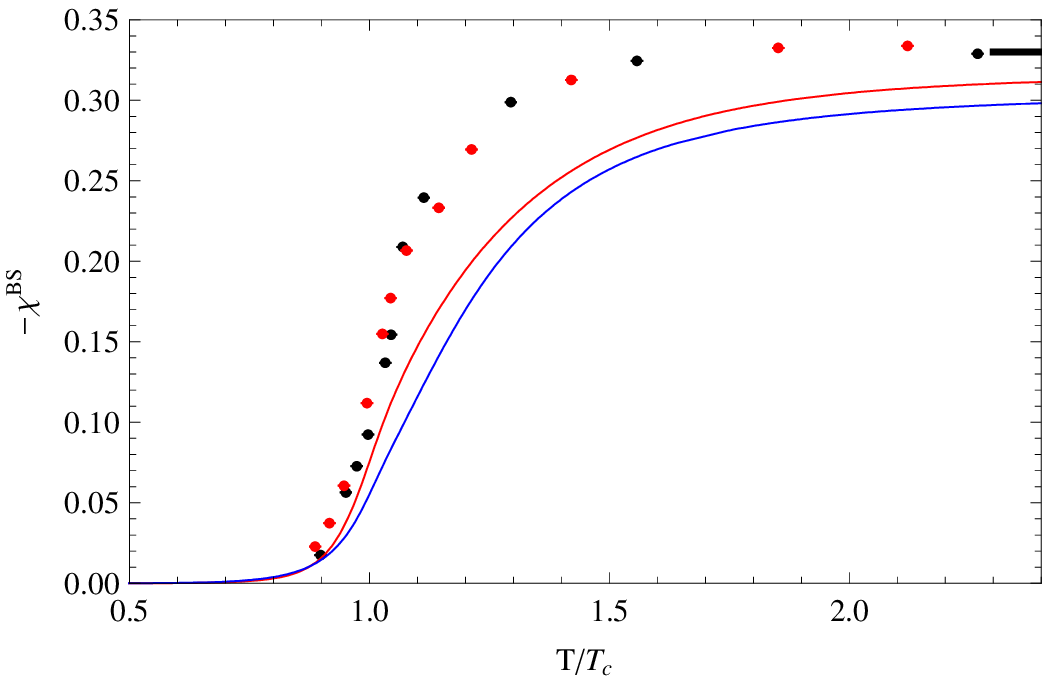}}
  \scalebox{0.75}{\includegraphics{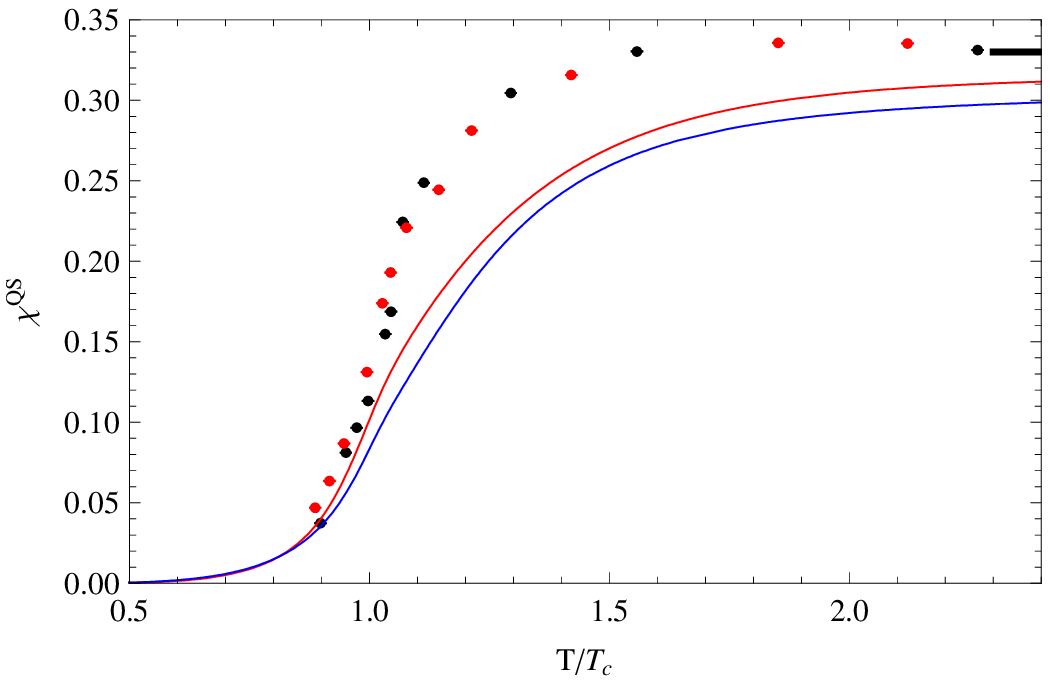}}
\caption{Plots of all the second order susceptibilities in ModelHotQCD (in red)
and ModelWB (in blue). The lattice data in black and red are obtained using p4 action with $N_{\tau}=4$ and $6$ respectively by the HotQCD collaboration~\cite{HotQCDlat2}. The blue lattice data are continuum estimate from WB collaboration using stout action~\cite{WBlat}. In case of $\chi_2^Q$, we have also plotted the WB lattice data for $N_{\tau}=8$(brown), 12(orange) and 16(green). In all the curves the high $T$ SB limit is indicated 
by a thick black line. }
\label{fg.chi2}
\end{figure}
All the second order susceptibilities, which measure fluctuation of the
conserved charges are small in the low $T$ regime and rise monotonically across the crossover 
temperature before saturating at high temperatures. This generic behaviour is seen both in the 
lattice data as well as in model calculations.
We have plotted all these quantities in Fig~\ref{fg.chi2}. The HotQCD data
for $\chi_2^B$, $\chi_2^Q$ and $\chi_2^S$ indicate a steep rise across the transition region
attaining the corresponding SB limits by $T\sim1.7$ $T_c$. On the other hand, the WB data rises
gently across $T_c$ attaining $90 \%$ of the SB limit by $T\sim2.4$ $T_c$. For $\chi_2^B$, ModelWB shows very
good agreement with the WB data. On the other hand, the rise in ModelHotQCD though faster than ModelWB, is not 
strong enough and fails to catch up with the HotQCD data. The ModelHotQCD reaches $93\%$ of the SB
limit by $T\sim2.4$ $T_c$. Similar high $T$ behaviour is observed for $\chi_2^Q$ and $\chi_2^S$. 
At low temperature, $\chi_2^Q$ from WB increases appreciably as $N_{\tau}$ is increased 
and the continuum limit is approached. This behaviour can be attributed to the fact that the dominant contribution to $\chi_2^Q$ comes from the pion. For the same reason, PQM at the mean field level fails to reproduce the WB data. Thus, it is necessary to incorporate pionic fluctuations in our present study for a better estimate of $\chi_2^Q$ (for similar studies in the context of the PNJL model, see~\cite{PNJL-fluc}).

The $\chi_{11}^{BQ}$ correlation exhibits a peak in the transition region. As already mentioned, in the 
low $T$ side because of the high mass of the relevant degrees of freedom, it goes to zero, while in the 
SB limit it is clear from (\ref{eq.muqtomuH}) that 
the contribution to $\chi_{11}^{BQ}$ from the $u$ quark sector is neatly cancelled by the sum of the
contributions from the $d$ and $s$ sectors and hence goes to zero again. ModelHotQCD agrees
reasonably well with the HotQCD data. In the peak region, it is about $40\%$ larger than
ModelWB. In case of the other correlators $\chi_{11}^{BS}$ and $\chi_{11}^{QS}$, the models compare 
in the same way to the lattice data as the diagonal susceptibilities.

\begin{figure}
  \scalebox{0.75}{\includegraphics{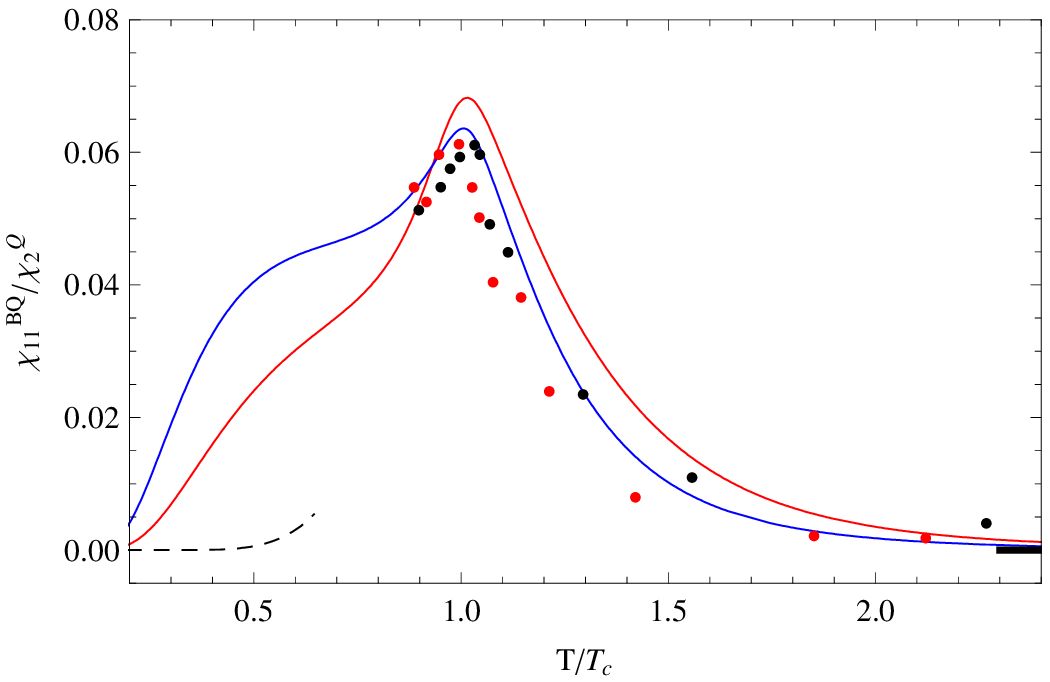}}
  \scalebox{0.75}{\includegraphics{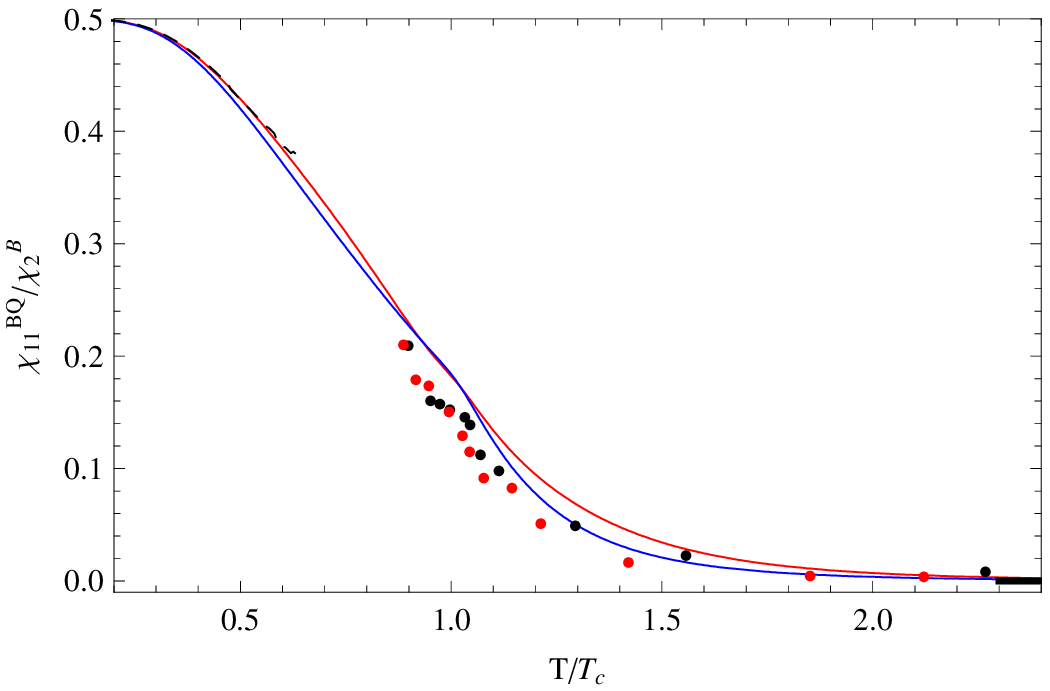}}\\
  \scalebox{0.75}{\includegraphics{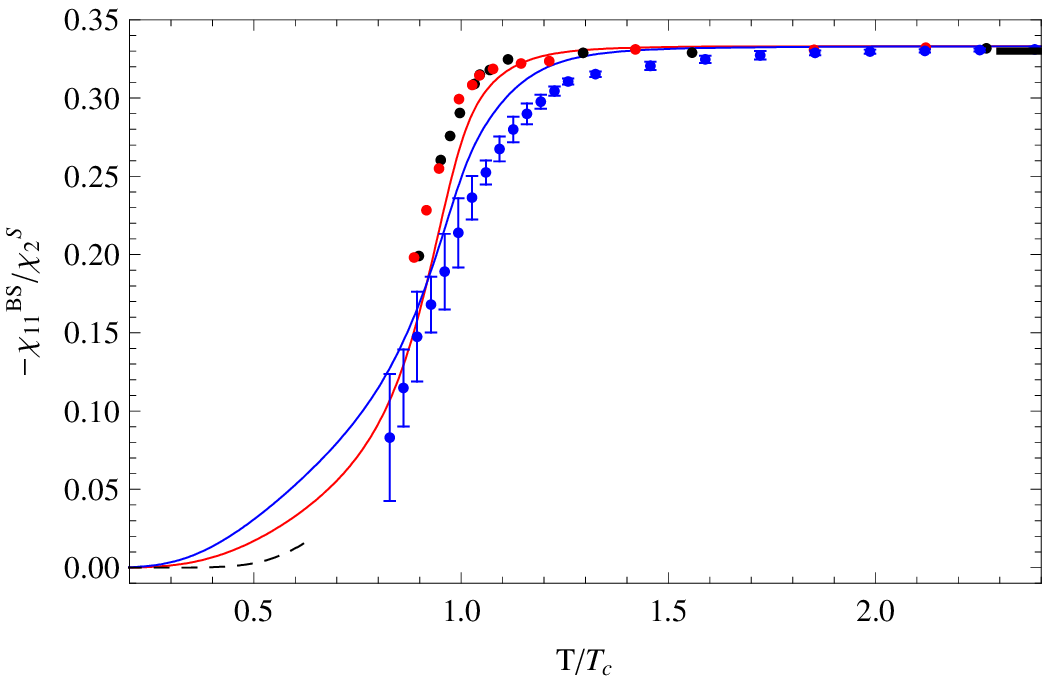}}
  \scalebox{0.75}{\includegraphics{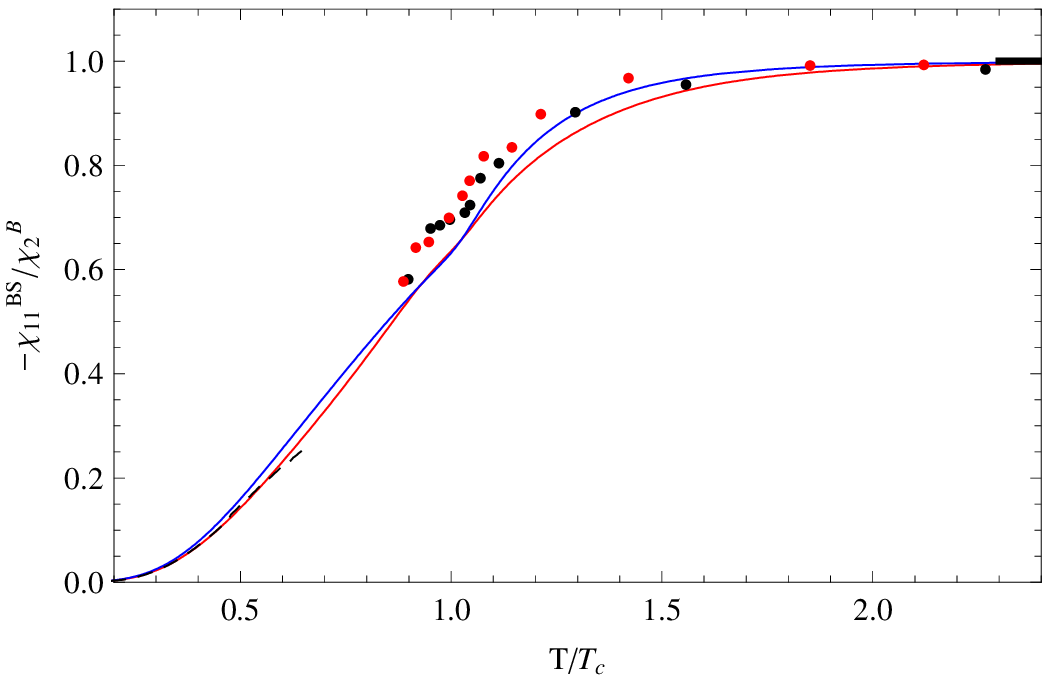}}\\
  \scalebox{0.75}{\includegraphics{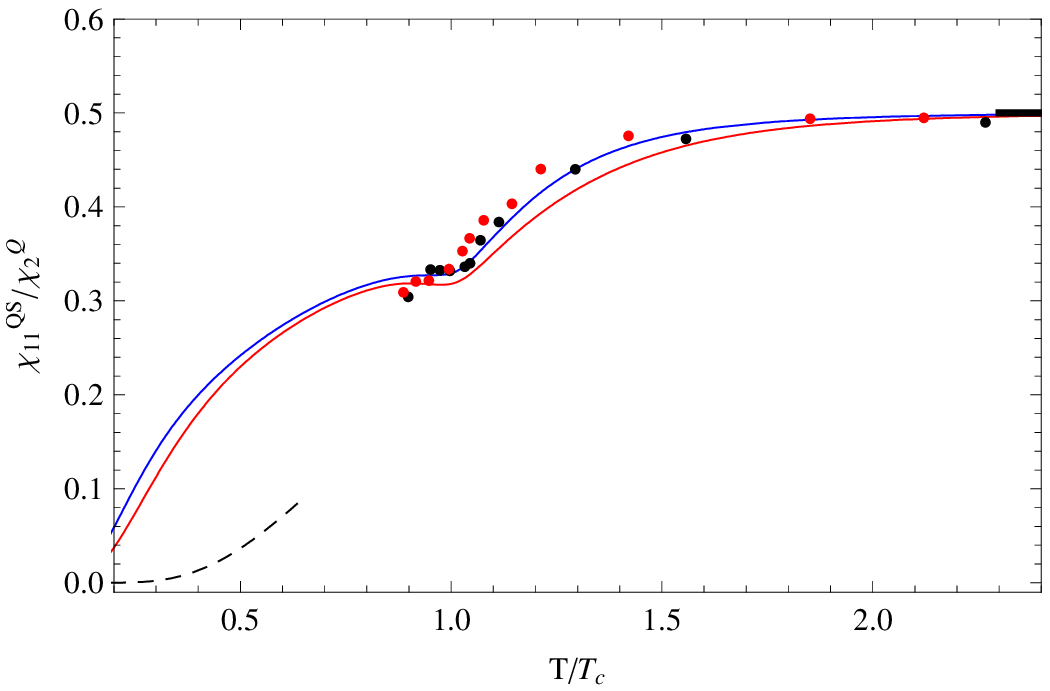}}
  \scalebox{0.75}{\includegraphics{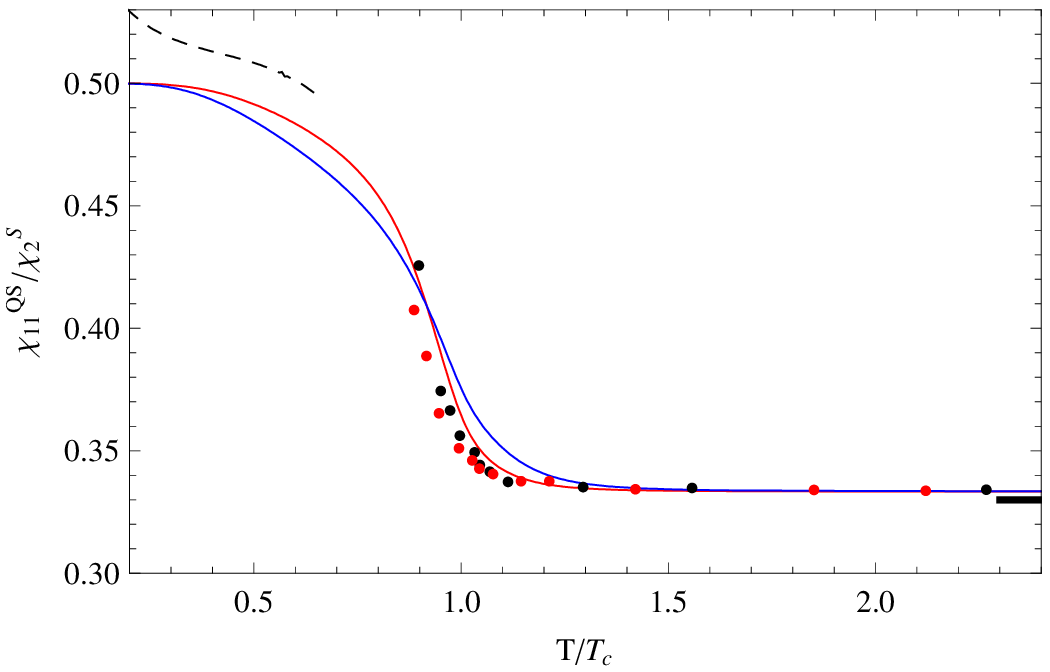}}
\caption{Plots of second order correlations normalised to the second order susceptibilities in 
ModelHotQCD (in red) and ModelWB (in blue). The lattice data in black and red are obtained using p4 action with $N_{\tau}=4$ and $6$ respectively by the HotQCD collaboration~\cite{HotQCDlat2}. The blue lattice data are continuum estimate from WB collaboration using stout action~\cite{WBlat}. In all the curves the high $T$ SB limit is indicated by a thick black line while the dashed black line at low $T$ shows the HRGM value.}
\label{fg.chi2r}
\end{figure}
We shall now focus on the ratio of the above susceptibilities. We have plotted these ratios in Fig. \ref{fg.chi2r}. The normalized correlations appear to show a better agreement with lattice data as compared to the unnormalized correlations themselves of Fig \ref{fg.chi2}. In these
plots, the ModelHotQCD and ModelWB lie almost together. $\chi_{11}^{BQ}/\chi_2^{Q}$ shows a peak in the transition region and both in the low and high $T$ regimes, this ratio goes to zero. In the high $T$ limit, $\chi_{11}^{BQ}$ itself goes to zero and thus the ratio goes to zero. In the low $T$ limit, the dominant contribution to $\chi_{11}^{BQ}$ is from the effective '3 quark state' from the non strange sector while the dominant contribution to $\chi_2^{Q}$ comes from the pion. The three quark state being much heavier than the pion, $\chi_{11}^{BQ}$ approaches zero much faster than $\chi_2^{Q}$ and thus the ratio goes to zero in the low $T$ limit. In the case of $\chi_{11}^{BQ}/\chi_2^{B}$, there is no peak like structure and the ratio decreases
monotonically. In this case the high $T$ behaviour is same as $\chi_{11}^{BQ}/\chi_2^{Q}$ and it goes to zero as $\chi_{11}^{BQ}$ itself goes to zero. In the low $T$ limit, both 
$\chi_{11}^{BQ}$ and $\chi_2^{B}$ get contribution from the '3 quark state' from the non strange sector. It is straightforward to check from (\ref{eq.muqtomuH}) that because of the relative sign between $\mu_B$ and $\mu_Q$ in the expression of $\mu_d$, $\chi_{11}^{BQ}/\chi_2^{B}$ approaches 1/2.
While $-\chi_{11}^{BS}/\chi_2^{S}$~\cite{C_BS} shows a steep rise and reaches the SB 
limit by $T/T_c\sim 1.3$, $-\chi_{11}^{BS}/\chi_2^{B}$ rises gently before meeting the SB limit at $T/T_c\sim 1.8$. In the low $T$ limit, we expect the ratios to go to zero: since the three quark states from the strange sector (which dominantly contribute to $\chi_{11}^{BS}$ in PQM) are much heavier compared to the three quark states from the non strange sector and the kaons which dominantly contribute to $\chi_2^{B}$ and $\chi_2^{S}$ respectively. As one can verify from Table~\ref{tb.SB}, $\chi_{11}^{QS}/\chi_2^{Q}$ and $\chi_{11}^{QS}/\chi_2^{S}$ approach 1/2 and 1/3 in the high $T$ limit by $1.8T_c$ and $1.1T_c$ respectively. Their low $T$ behaviour is quite distinct. Since the kaons which contribute dominantly to $\chi_{11}^{QS}$ are much heavier than the pions that has a leading contribution to $\chi_{2}^{Q}$, $\chi_{11}^{QS}/\chi_2^{Q}$ goes to zero. On the other hand, kaons alone contribute dominantly to both $\chi_{11}^{QS}$ and $\chi_{11}^{S}$ and the ratio approaches 1/2 as only two out of the four kaons are charged. $\chi_{11}^{QS}/\chi_2^{Q}$ also exhibits interesting feature in the transition regime, a plateau structure near $T_c$. Similar observation has been found even on the lattice~\cite{HotQCDlat2} as well as in the case of PNJL~\cite{PNJLsus_china2,PNJLsus_kol2} model. In~\cite{PNJLsus_kol2} this has been attributed to the shift in the role of the dominant degrees of freedom from the hadrons to quark quasiparticles just above $T_c$.

At low temperatures, as argued earlier the ratios obtained in PQM are found to compare well with that of HRGM which are also shown for comparison. The correlator $\chi_{11}^{BQ}$
project only on the charged baryon sector. Thus within HRGM at low temperatures, the ratio $\chi_{11}^{BQ}/\chi_2^{B}$ approach $\frac{1}{2}$ as $\chi_2^{B}$ gets contribution from both
protons and neutrons while $\chi_{11}^{BQ}$ gets contribution only from protons. For similar reasons, $\chi_{11}^{QS}/\chi_2^S$ also approach 
$\frac{1}{2}$. We note that the HRGM plot of $\chi_{11}^{QS}/\chi_2^S$ in Fig.~\ref{fg.chi2r} shows a value slightly greater
than $0.5$ which is due to the small mass difference between the different isospin states of kaon (which is the lightest relevant hadron in 
this case) . $\chi_{11}^{BS}/\chi_2^{B}$ go to zero at low $T$ because the lightest baryons do not carry strangeness. Similarly, 
$\chi_{11}^{BQ}/\chi_2^{Q}$, $\chi_{11}^{BS}/\chi_2^{S}$, $\chi_{11}^{BS}/\chi_2^{B}$ and $\chi_{11}^{QS}/\chi_2^{Q}$ go to zero too.

Thus we see that all the second order fluctuations increase monotonically across $T_c$.
In the case of the correlations, the story is more varied: while $\chi_{BQ}$ exhibits a peak in the transition regime, $\chi_{BS}$ and $\chi_{QS}$ rise monotonically across $T_c$. When we consider the ratios, we have an even more varied features. $\chi_{11}^{BQ}/\chi_2^B$ and $-\chi_{11}^{BS}/\chi_2^B$ show a monotonic beahviour, 
$\chi_{11}^{BS}/\chi_2^{S}$ and $\chi_{11}^{BS}/\chi_2^{B}$ exhibit sharp changes in the transition regime, $\chi_{11}^{BQ}/\chi_2^Q$ shows a sharp peak and $\chi_{11}^{QS}/\chi_2^Q$ exhibit a plateau in the transition regime. Thus quantities like $\chi_{11}^{BQ}$, $\chi_{11}^{BQ}/\chi_2^Q$, $\chi_{11}^{BS}/\chi_2^{S}$, $\chi_{11}^{BS}/\chi_2^{B}$ and $\chi_{11}^{QS}/\chi_2^Q$ are well suited to probe the QCD phase transition because of their distinct features in the transition regime. In fact, it was already pointed out in~\cite{C_BS} that the ratio $\chi_{11}^{BS}/\chi_2^{S}$ has different values in the hadron and QGP phases and thus is a very sensitive tool to probe the nature of the strongly interacting matter.
\subsection{$\l i+j+k \r = 4$}
\begin{figure}
  \scalebox{0.75}{\includegraphics{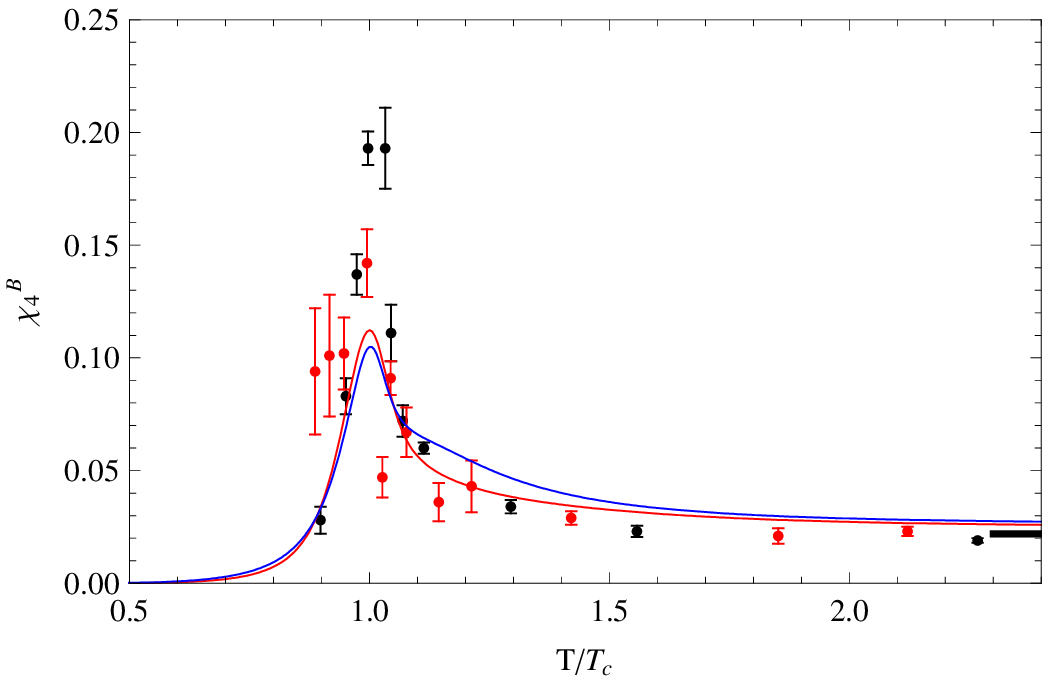}}
  \scalebox{0.75}{\includegraphics{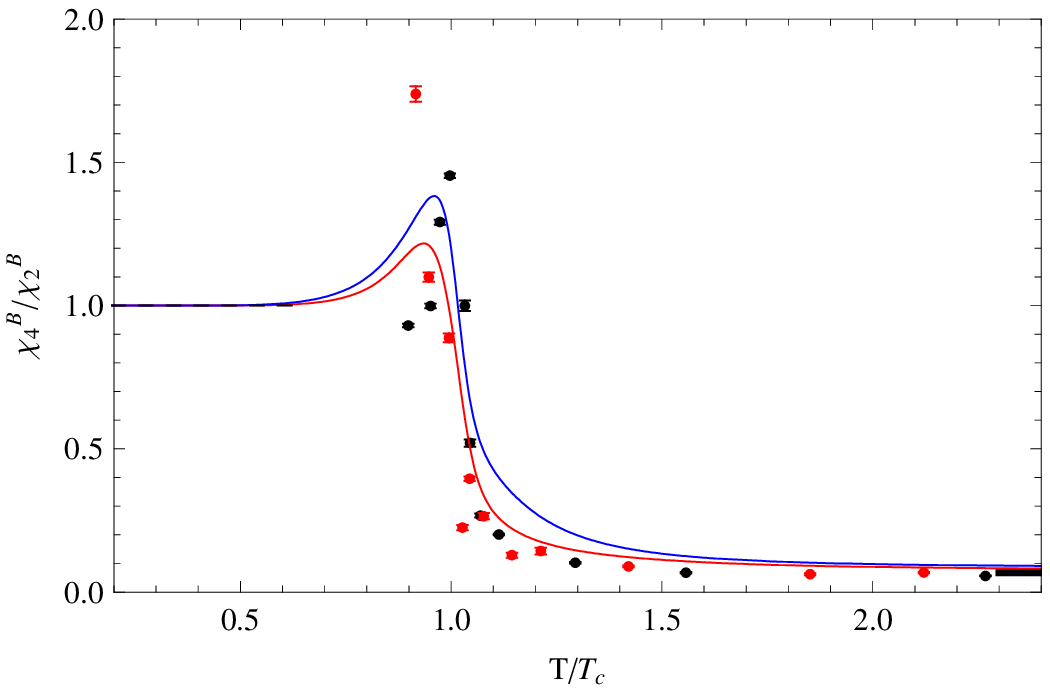}}\\
  \scalebox{0.75}{\includegraphics{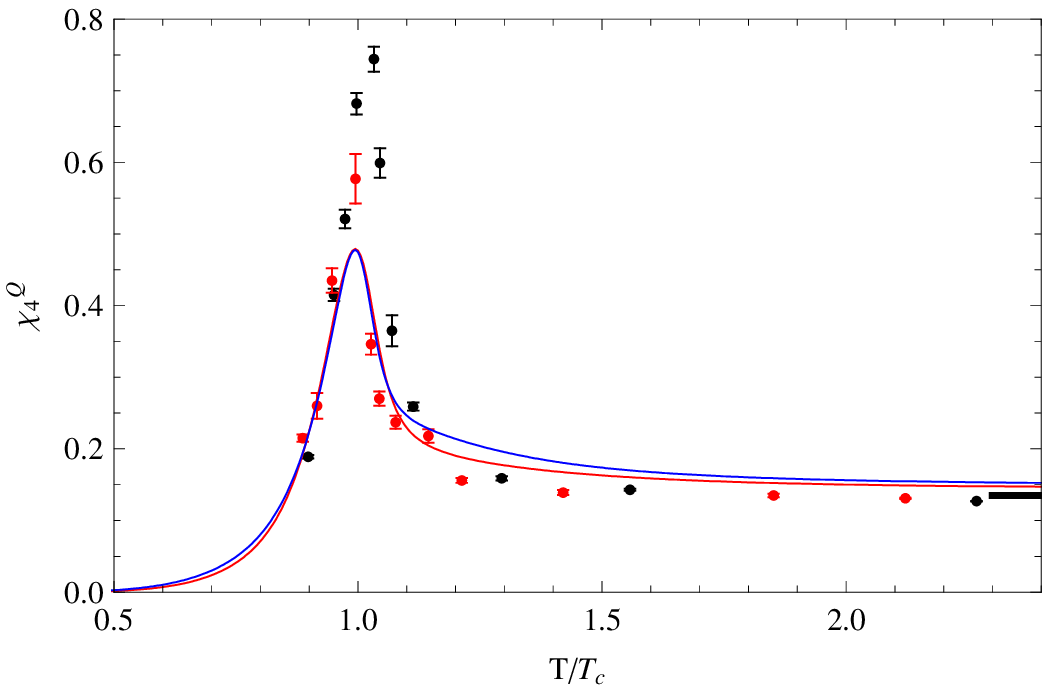}}
  \scalebox{0.75}{\includegraphics{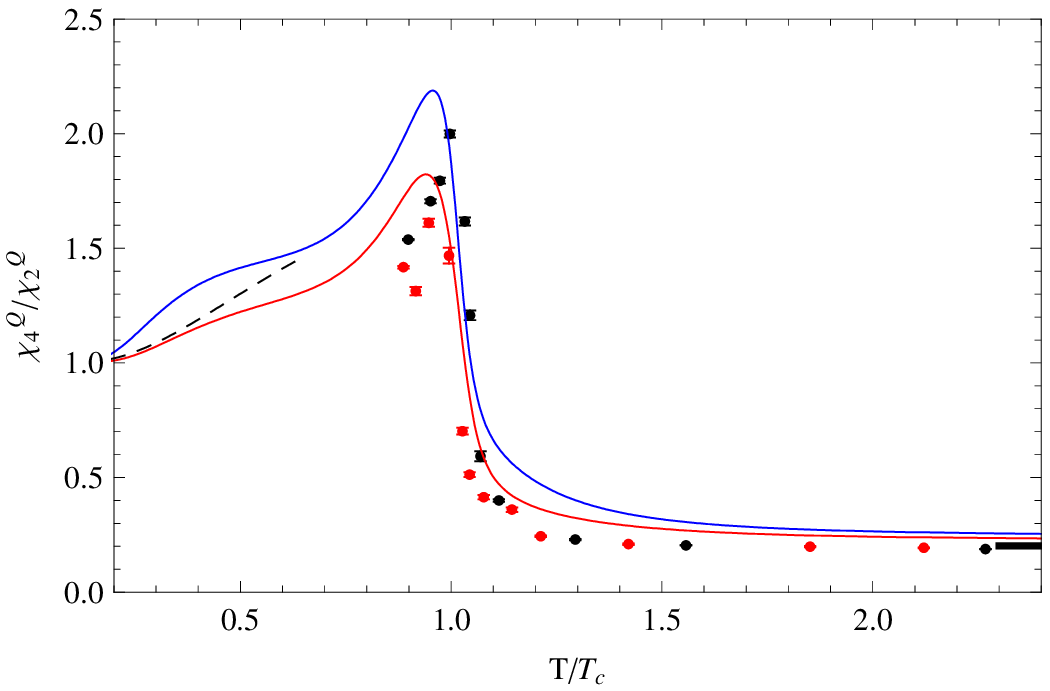}}\\
  \scalebox{0.75}{\includegraphics{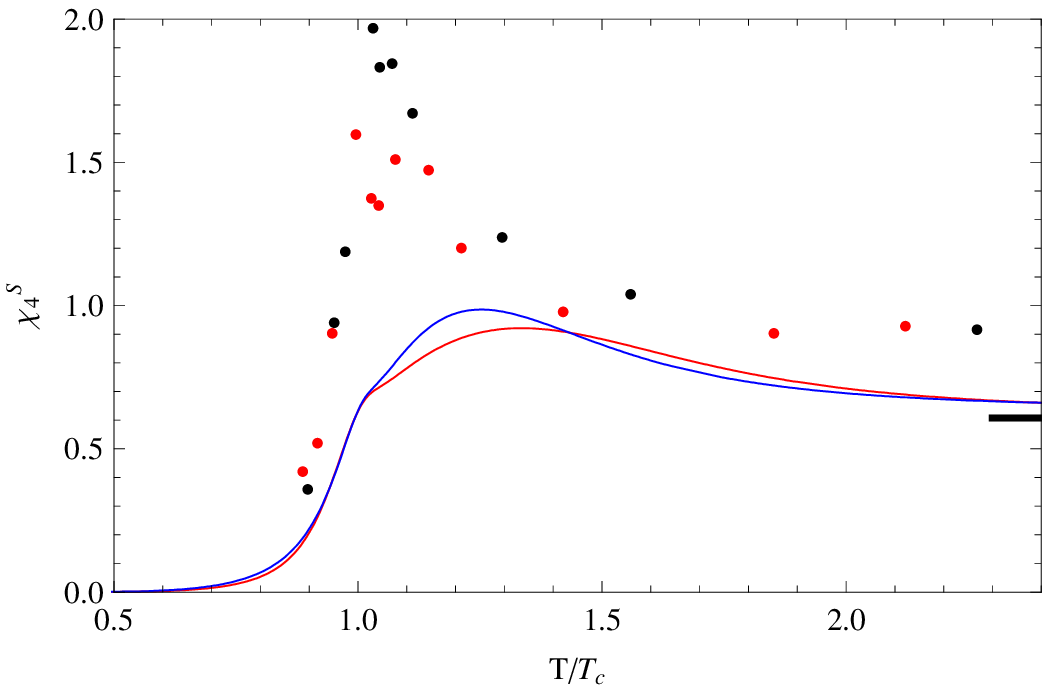}}
  \scalebox{0.75}{\includegraphics{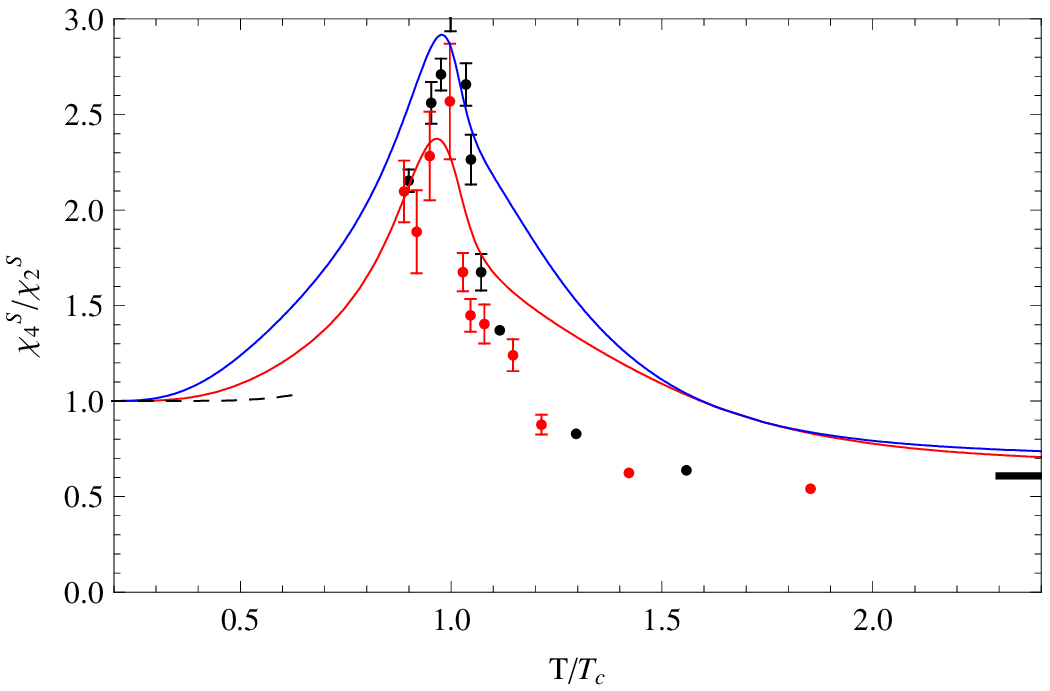}}
\caption{On the left are plots of fourth order susceptibilities and on the right are plots of 
these susceptibilities normalized by their second order susceptibilities in ModelHotQCD (in red) 
and ModelWB (in blue). The lattice data in black and red are obtained using p4 action with $N_{\tau}=4$ and $6$ respectively by the HotQCD collaboration~\cite{HotQCDlat2}. In all the curves the high $T$ SB limit is indicated 
by a thick black line while the dashed black line at low $T$ shows the HRGM value.}
\label{fg.chi4}
\end{figure}

The higher order susceptibilities are more 
sensitive to fluctuations in a system. We show in the left column of Fig \ref{fg.chi4} plots of fourth order diagonal 
susceptibilities $\chi^4_B$, $\chi^4_Q$ and $\chi^4_S$. Unlike the second order diagonal susceptibilities,
in this case ModelHotQCD and ModelWB almost lie on top of each for the entire temperature range.
A notable feature for all the three susceptibilities is that they all peak around the cross-over temeperature
which is also evident from LQCD. However, when compared to lattice the models show a smaller peak. 
But here we should note that the data for $N_{\tau}=6$ are consistently smaller than the $N_{\tau}=4$ data
and so one would expect that the continuum estimates of LQCD will further come down which will be in
better agreement with the PQM predictions. $\chi^4_S$ shows a very gentle peak. This is a common 
feature of effective models (also seen in PNJL~\cite{PNJLsus_china2} models), where the strange 
quark mass reduces at a slower rate as compared to the lighter quarks in the transition region.
In the high $T$ region, both model as well as LQCD seem to attain the SB value in the case of $\chi^4_B$ 
and $\chi^4_Q$ while for $\chi^4_S$ the models are $11\%$ higher than 
the SB limit, whereas lattice is $50\%$ higher at $T/T_c\sim 2.4$.

In the right column of Fig \ref{fg.chi4}, we have shown plots of the ratio of the fourth order
susceptibilities to those of the corresponding second order ones. As in the case of second order correlations, 
it is worthwhile to note that the models are in better agreement with lattice studies when the ratios 
of the susceptibilities are compared. In all the cases, ModelHotQCD predictions are slightly lower as
compared to that of ModelWB. As argued earlier, at low $T$, ratios of the kind $\chi^X_i/\chi^X_j$ where $X \in B, Q, S$ approach unity both in PQM as well as in HRGM. In the crossover region, all the ratios exhibit a peak, the most prominent being that of $\chi_4^B/\chi_2^B$ which 
starts deviating from unity only around $T \sim 0.8 T_c$ making it a good candidate to probe the crossover 
region~\cite{chi4Btochi2B}. In the high $T$ limit, the model predictions saturate around the SB values.

\begin{figure}
  \scalebox{0.5}{\includegraphics{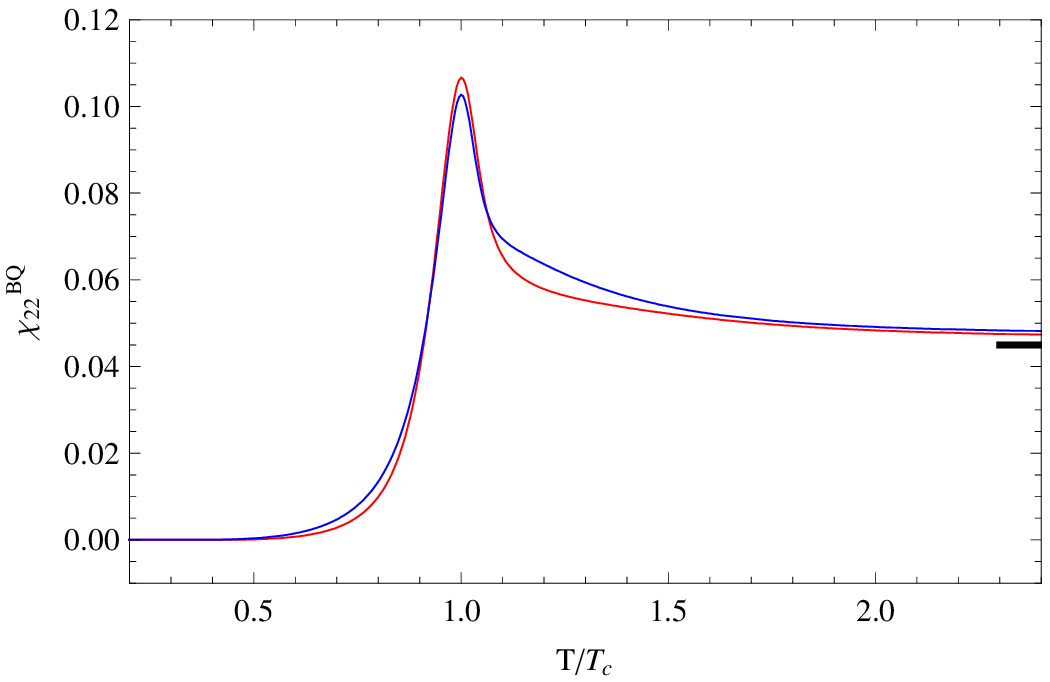}}
  \scalebox{0.5}{\includegraphics{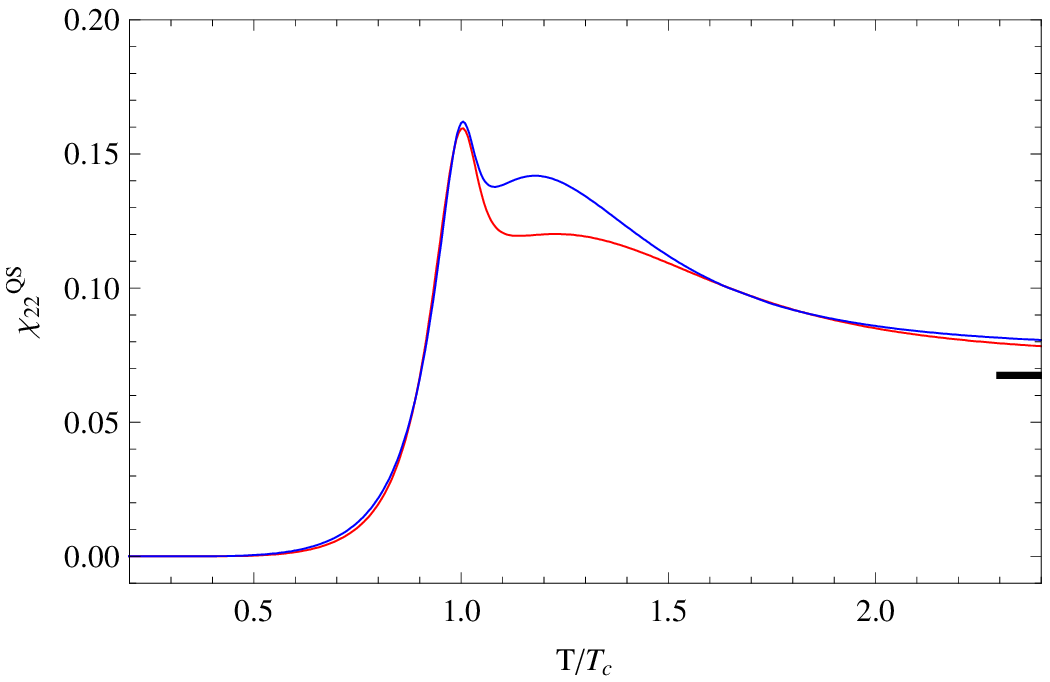}}
  \scalebox{0.5}{\includegraphics{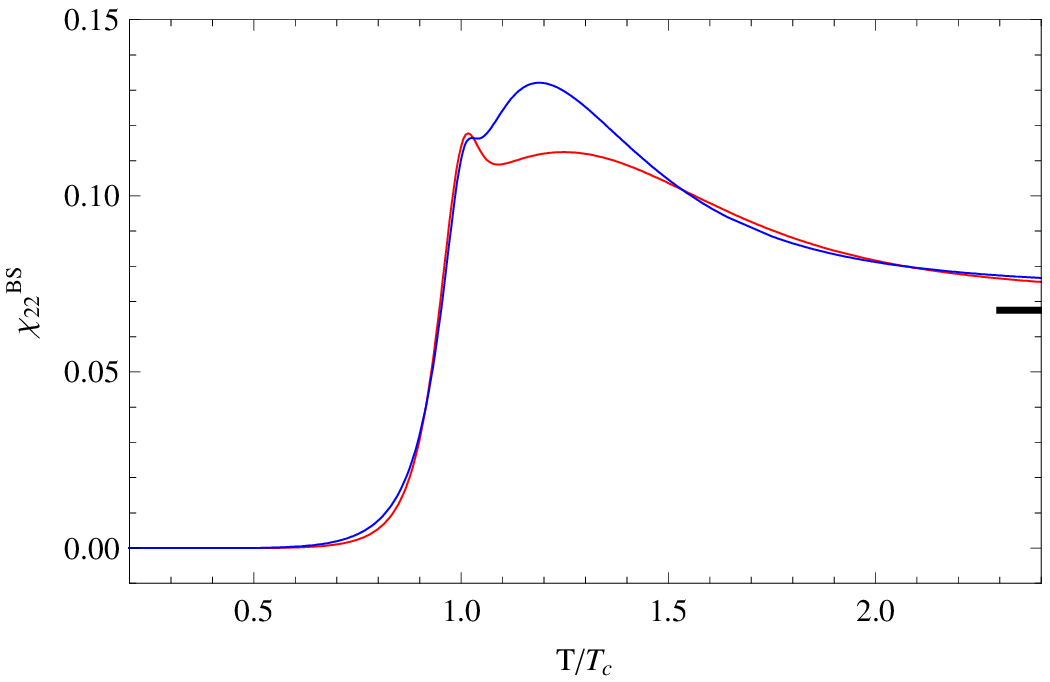}}\\
  \scalebox{0.5}{\includegraphics{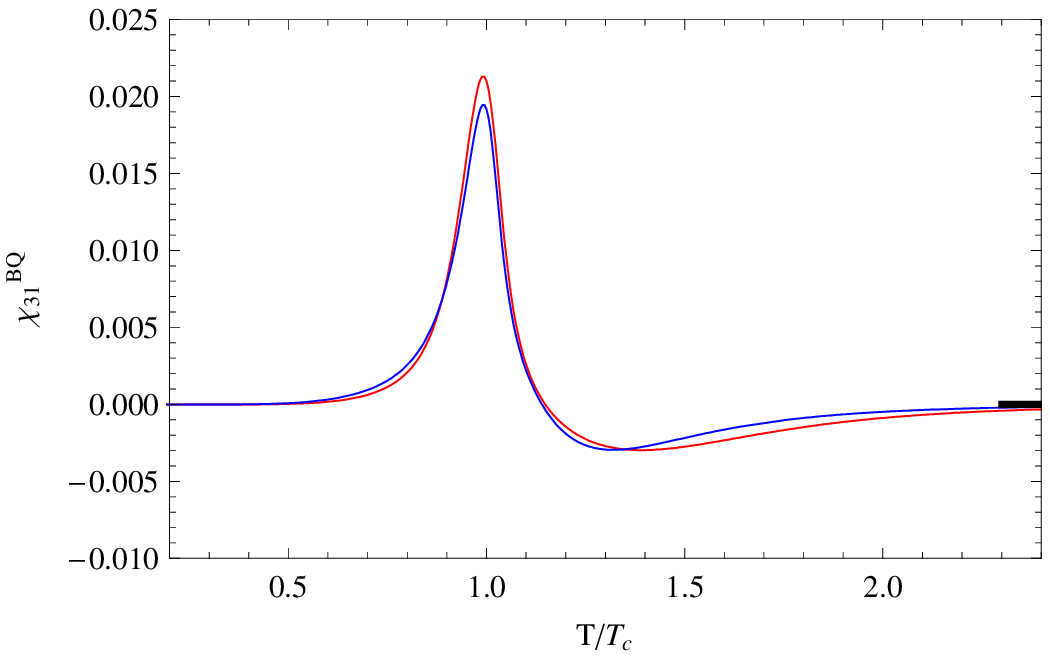}}
  \scalebox{0.5}{\includegraphics{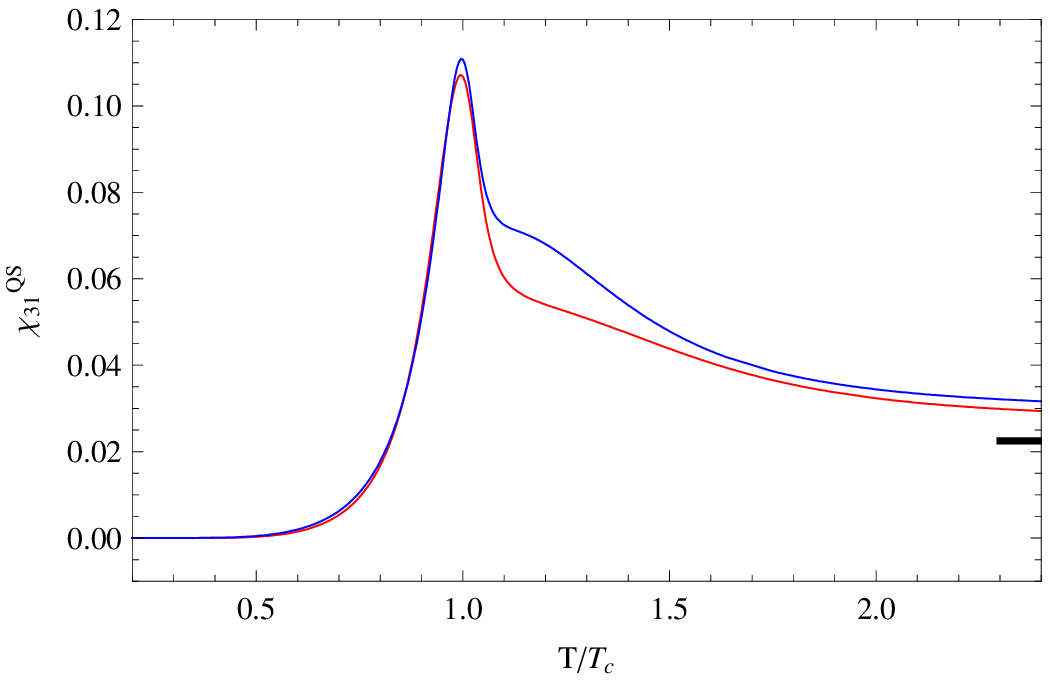}}
  \scalebox{0.5}{\includegraphics{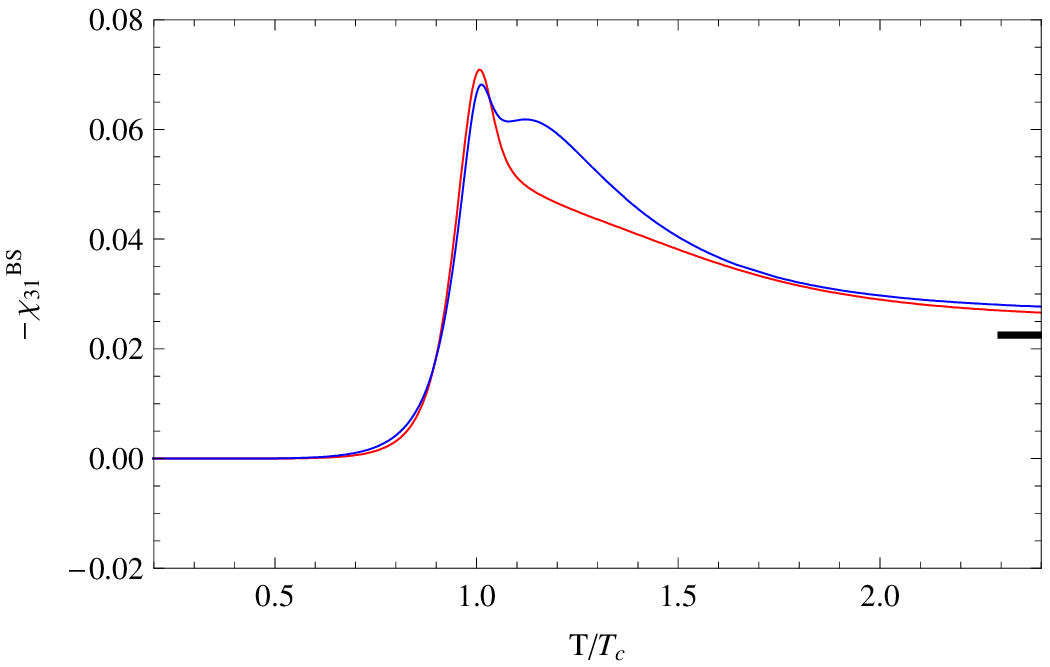}}\\
  \scalebox{0.5}{\includegraphics{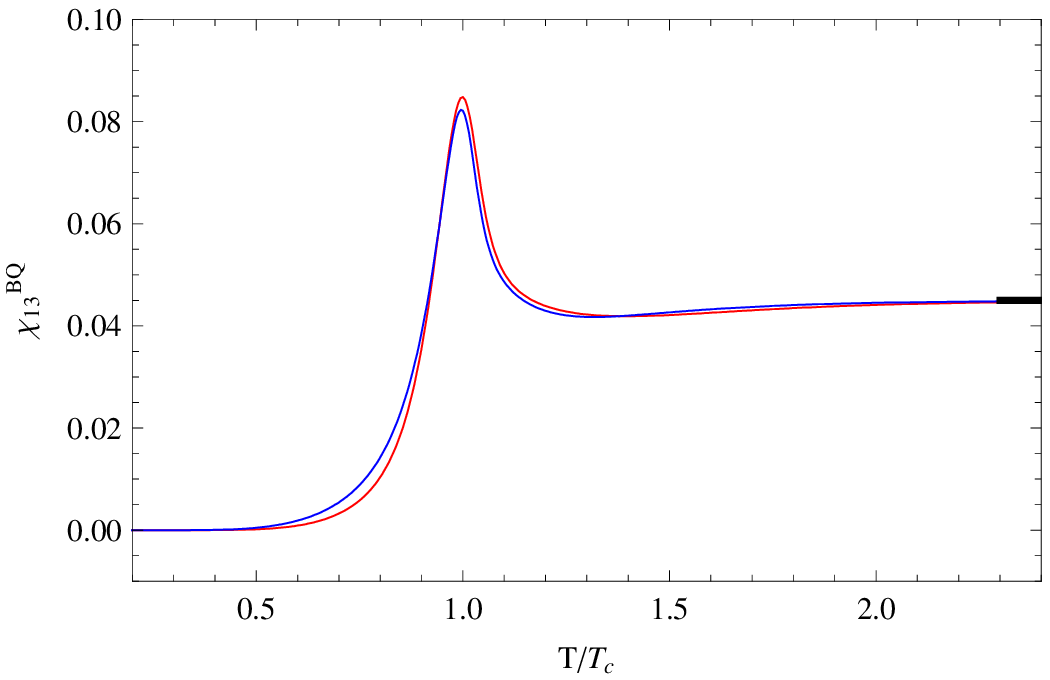}}
  \scalebox{0.5}{\includegraphics{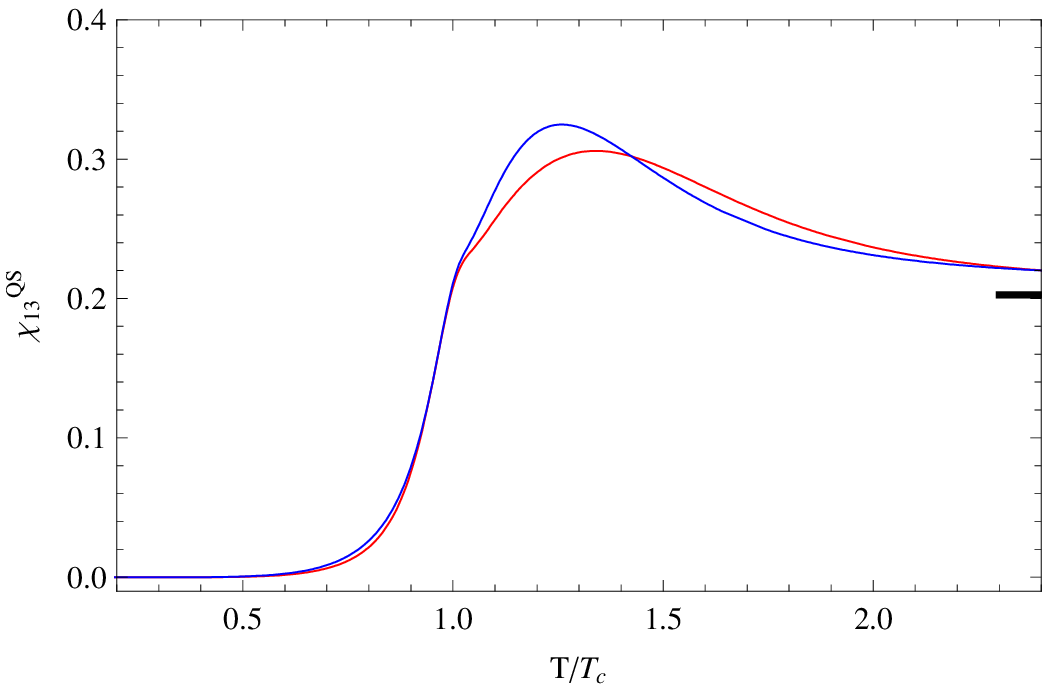}}
  \scalebox{0.5}{\includegraphics{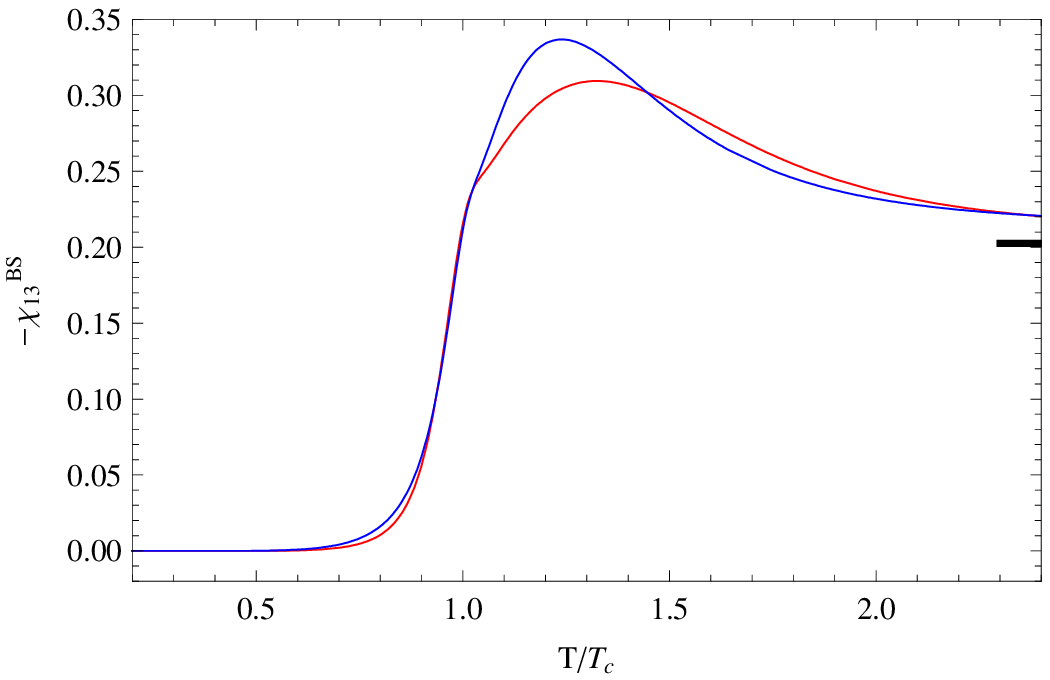}}\\
  \scalebox{0.5}{\includegraphics{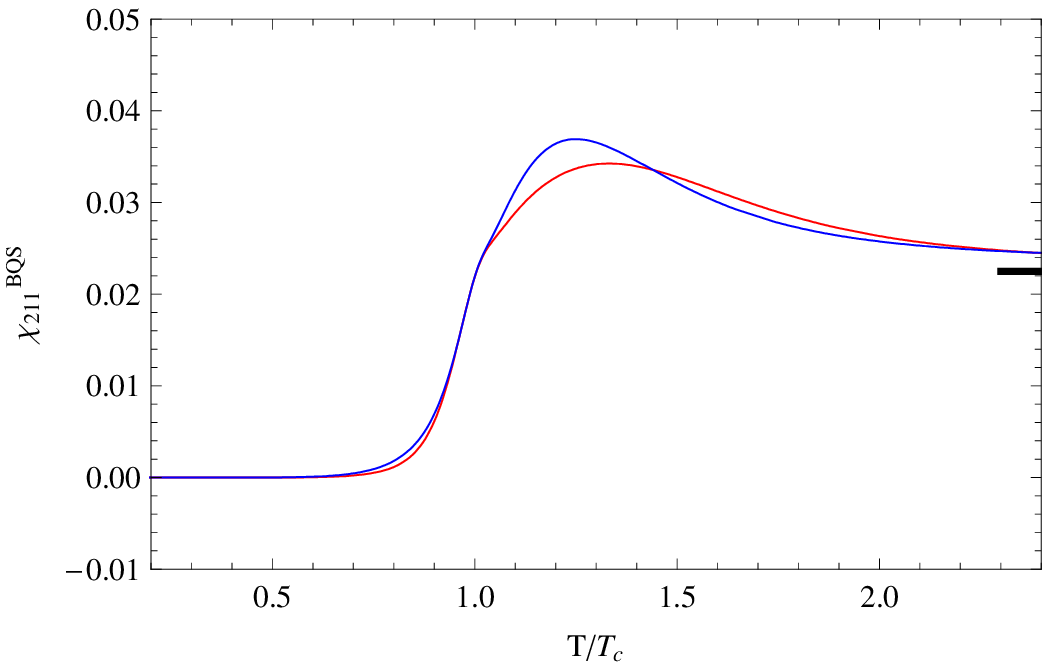}}
  \scalebox{0.5}{\includegraphics{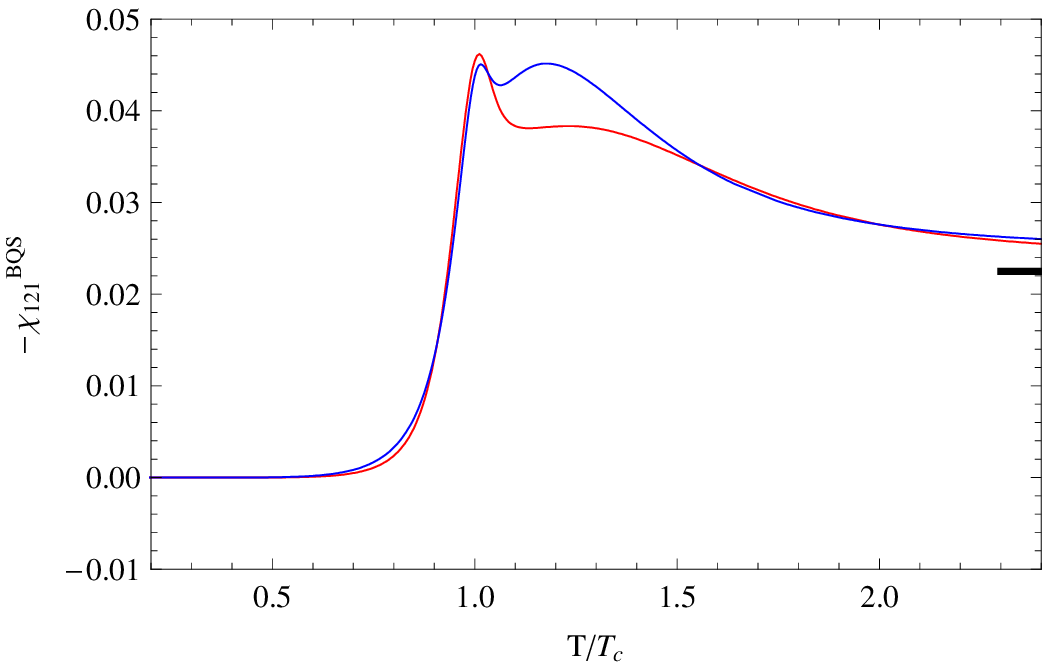}}
  \scalebox{0.5}{\includegraphics{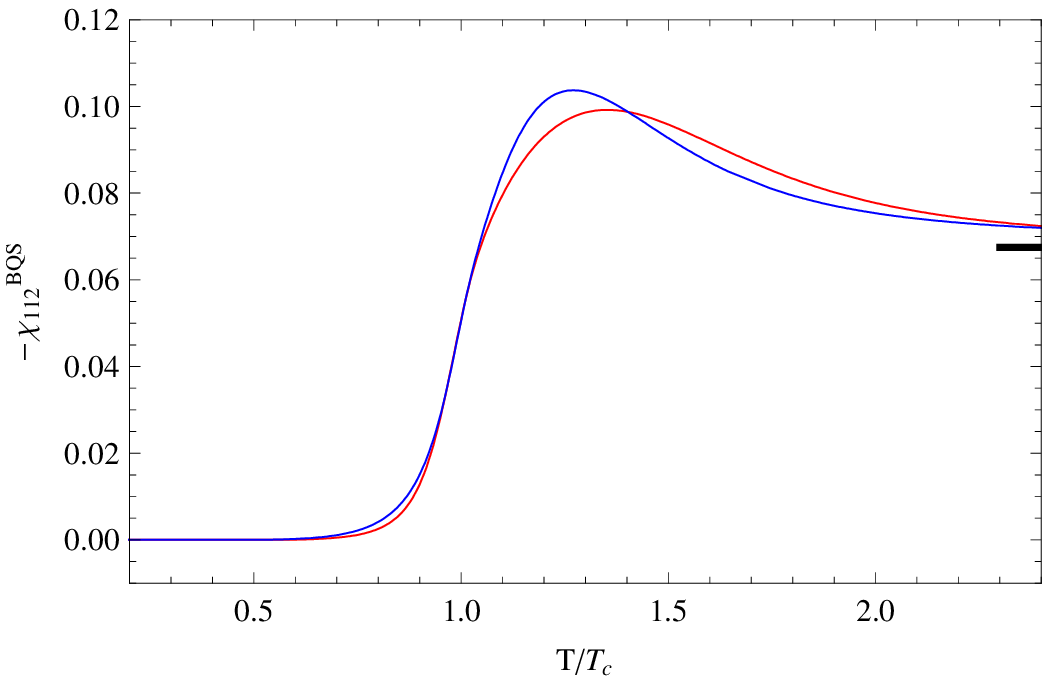}}
\caption{Fourth order correlations for ModelHotQCD (red) and ModelWB (blue) are plotted. In all the curves the high $T$ SB limit is indicated 
by a thick black line.}
\label{fg.chi4corr} 
\end{figure}

For completeness in Fig \ref{fg.chi4corr}, we have presented the model computation for all the fourth order correlations. There are total 16 of them. LQCD data for these correlations are still not available.
All the fourth order correlations stay at zero for temperatures upto $\sim 0.7 T_c$, while in the high $T$ limit they all approach within $10\%$ of the SB values. All the correlations in the BQ sector exhibit a sharp peak in the transition regime, while the introduction of strangeness broadens the peak in the other cases due to the slow melting of the strange condensate. For some cases like $\chi_{22}^{QS}$, $\chi_{31}^{QS}$, $\chi_{22}^{BS}$, $-\chi_{31}^{BS}$ and $-\chi_{121}^{BQS}$ there is a hint of a broad double peak in ModelWB, although in case of ModelHotQCD the second peak looks sufficiently suppressed. Similar double peak structures have also been seen in the case of $\l2+1\r$ PNJL model~\cite{PNJLsus_kol2}.

\subsection{$\l i+j+k \r = 6$}
\begin{figure}
 \scalebox{0.75}{\includegraphics{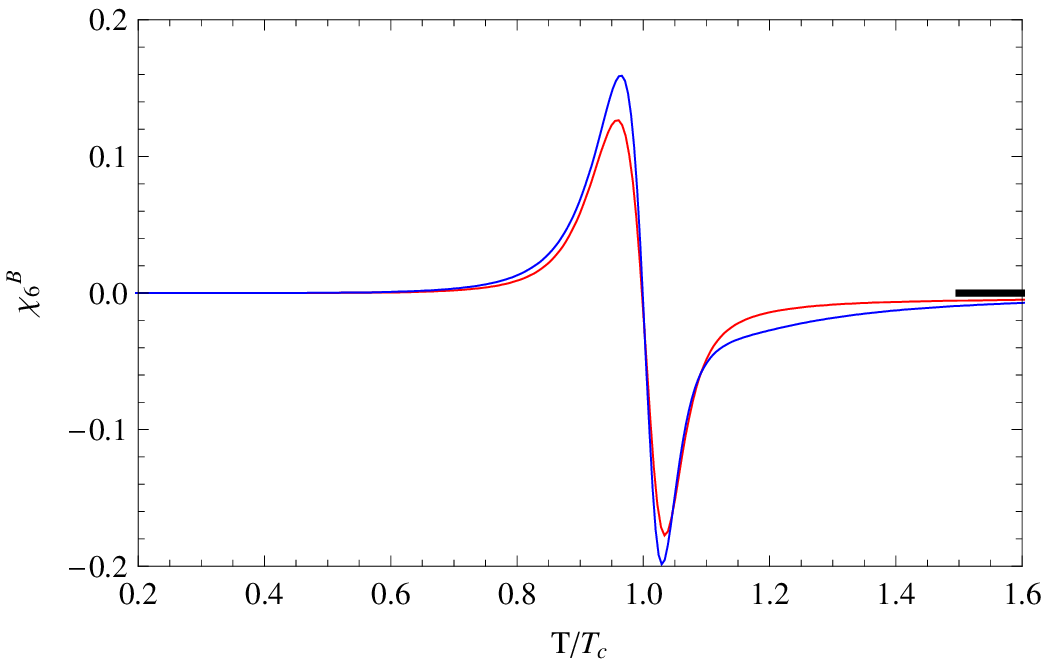}}
 \scalebox{0.75}{\includegraphics{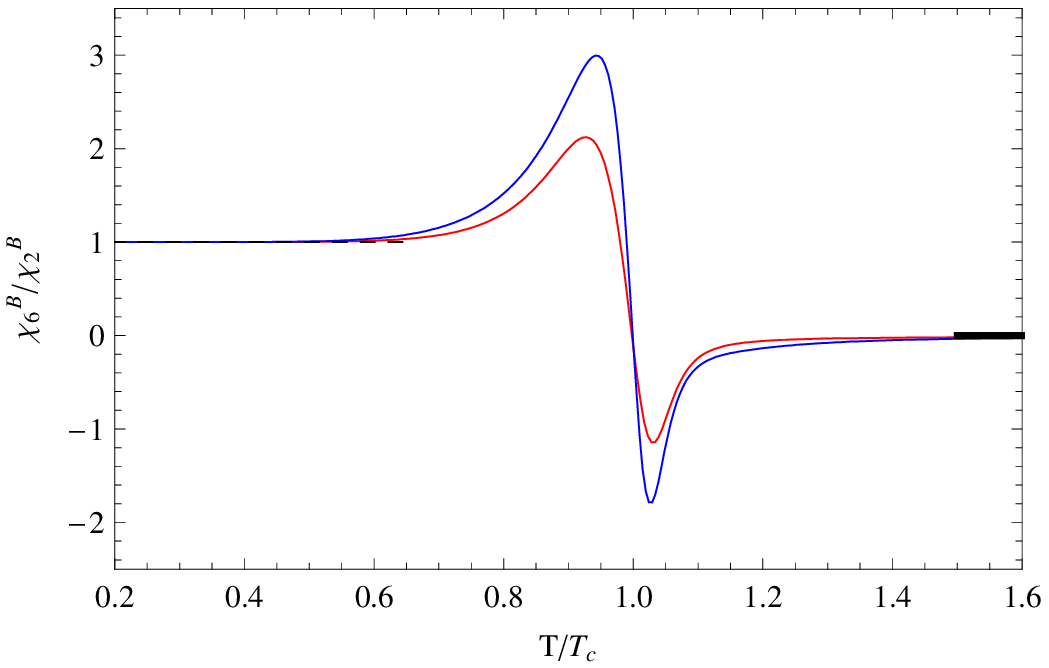}}\\
 \scalebox{0.75}{\includegraphics{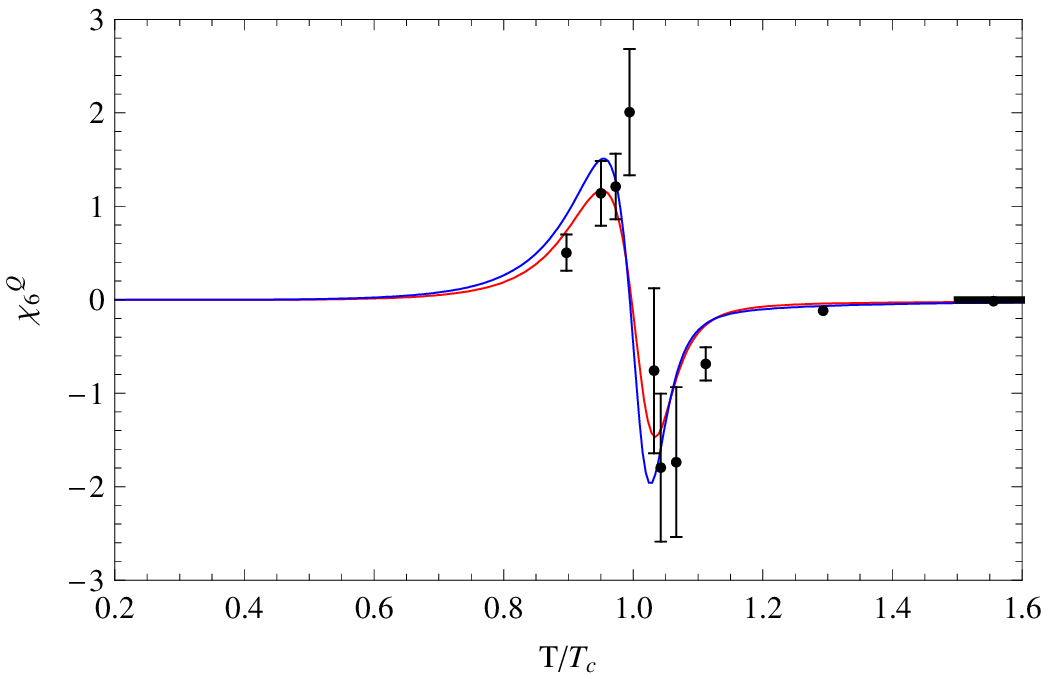}}
 \scalebox{0.75}{\includegraphics{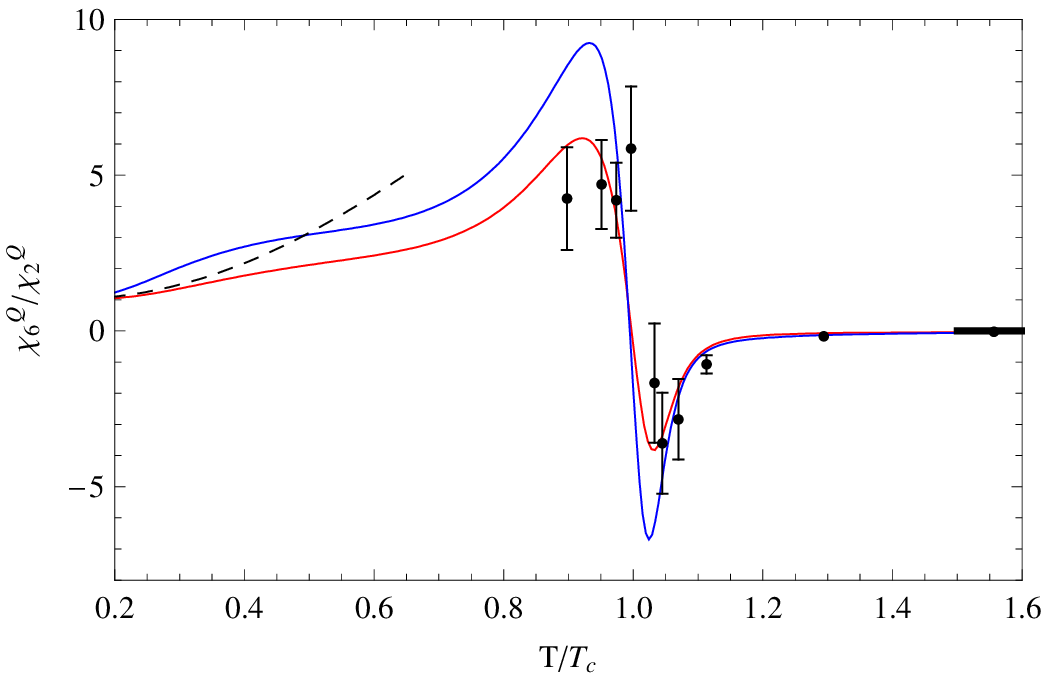}}\\
 \scalebox{0.75}{\includegraphics{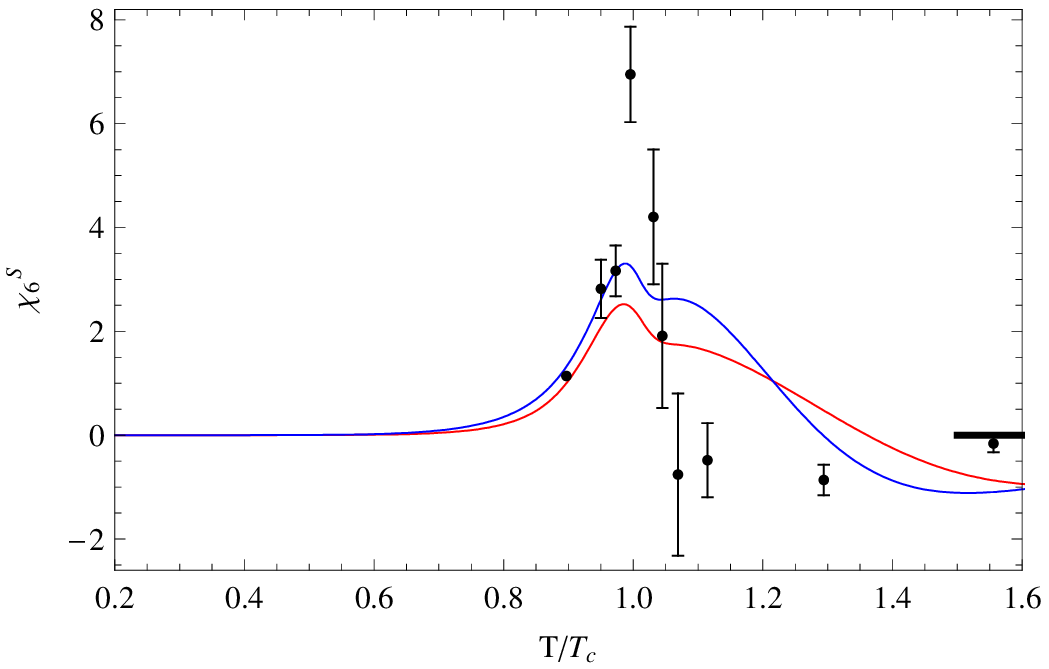}}
 \scalebox{0.75}{\includegraphics{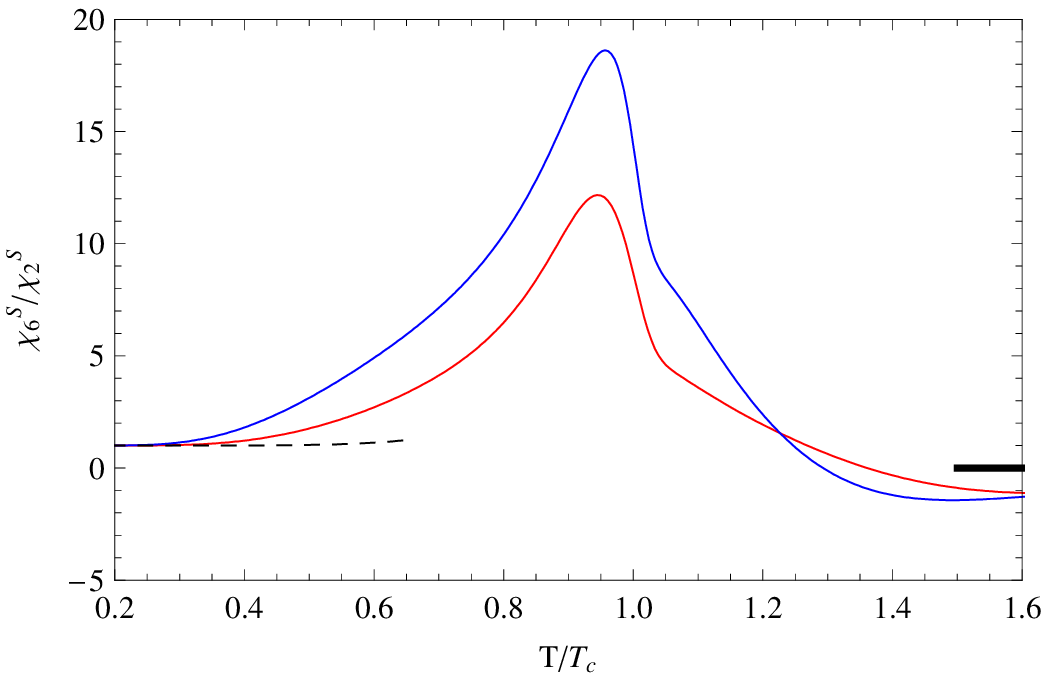}}
\caption{On the left are plots of sixth order susceptibilities and on the right are plots of these 
susceptibilities normalized by their second order susceptibilities in ModelHotQCD (in red) and 
ModelWB (in blue). The lattice data in black are obtained using p4 action with $N_{\tau}=4$ by the HotQCD collaboration~\cite{HotQCDlat2}. In all the curves the high $T$ SB limit is indicated by a thick black line while the dashed black line at low $T$ shows the HRGM value.}
\label{fg.chi6}
\end{figure}
There are LQCD data for some of the sixth order fluctuations as well. 
We have also computed the sixth order diagonal susceptibilities in PQM shown in Fig \ref{fg.chi6}. 
The highest power of the chemical potential that appears in the SB formula for pressure is four 
(see (\ref{eq.pSB})). Thus in the high temperature SB limit the sixth order susceptibilities go 
to zero. In general, these susceptibilities are very sensitive to the transition region, particularly $\chi_6^B$ and $\chi_6^Q$ which stay non-zero only in the temperature range from $0.8 T_c$ to $1.1 T_c$. Unlike the case of the fourth order diagonal susceptibilities where  all of them exhibited a 
peak in the crossover region, the sixth order ones oscillate between positive and negative values 
passing through zero once in the transition region. Such oscillatory behaviour has also been noted in 
the case of PNJL studies~\cite{PNJLsus_china1,PNJLsus_kol1}. We have also plotted the normalised
susceptibilities by taking ratios with respect to the second order ones. As expected, these ratios approach unity in the low $T$ limit both in PQM and HRGM. While $\chi_6^B/\chi_2^B$ stays at unity upto almost $0.8 T_c$, $\chi_6^Q/\chi_2^Q$ and $\chi_6^S/\chi_2^S$ show deviations from unity earlier. Once again the PQM model
predictions for the ratios are found to interpolate well between the HRGM results for low $T$ and those of an ideal gas of three massless quarks in the high temperature limit.

At zero chemical potential in the chiral limit, the singular behaviour of the quadratic and higher order baryon number fluctuations are supposed to be controlled by the $O\l4\r$ symmetry group with the scaling behaviour~\cite{baryonscaling}
\beq
\chi_{2n}^B\sim\left|\frac{T-T_c}{T_c}\right|^{2-n-\alpha}
\label{eq.scale}
\eeq
where $\alpha\simeq-0.25$. Thus with higher order, the singular property of the baryon number fluctuations in the vicinity of $T_c$ grow. Similar behaviour is also expected for elctric charge fluctuations~\cite{electricscaling}. Our study with PQM for $\l2+1\r$ flavors with physical light quark masses show that the behaviour of the baryon number fluctuations in the transition region is still controlled by (\ref{eq.scale}) to a certain extent as also seen in LQCD. While the quadratic fluctuations grow monotonically across the transition region, the quartic fluctuations seem to show a cusp like behaviour in the crossover regime. The singular nature grows even stronger in the case of the sixth order fluctuations and we find oscillatory behaviour.

\section{Summary and Conclusion}
\label{conclusion}
We shall now summarise our present study. In our earlier work~\cite{PQMVT-2+1}, we had studied the influence of the frequently omitted fermionic vacuum fluctuations on the thermodynamics of the $\l2+1\r$ flavor PQM and showed that the inclusion of this vacuum term leads to considerable improvement of the comparison between the model and LQCD. In this study we continue our investigation of the vacuum term in the PQM model and focus particularly on the generalised susceptibilities of the conserved charges B, Q and S upto order $6$ for two different parameter sets of the model as 
obtained earlier~\cite{PQMVT-2+1}, ModelHotQCD and ModelWB and compare our results with those of LQCD from HotQCD and WB groups. In this study we restrict our investigation to zero chemical potentials where LQCD data is currently available. 

The second order diagonal susceptibilities for $B$ and $S$ of ModelWB are in excellent agreement with WB lattice data, whereas those of ModelHotQCD fall short by $\sim7\%$ as compared to the HotQCD data in the high $T$ regime. In case of $\chi_2^Q$, although in the high $T$ limit there is good agreement with LQCD, in the low $T$ limit, the ModelWB falls considerably short to that of the WB continuum estimate. This can be attributed to the fact that pionic fluctuations contribute dominantly to charge fluctuations in the low $T$ limit which can not be captured presently in this mean field level analysis. The shape of the second order correlations nicely agree with that of LQCD: the $BQ$ correlator shows a peak, although ModelHotQCD shows an enhanced peak by about $40\%$ as compared to that of ModelWB. On the other hand, both $\chi_{11}^{QS}$ and $-\chi_{11}^{BS}$ rises monotonically across the transition regime, with the model predictions falling short by $\sim30\%$ to the HotQCD data in the transition region.

The behaviour of the baryon number fluctuations in the transition regime are controlled by the corresponding singular behaviour (\ref{eq.scale}) in the chiral limit. While the second order diagonal susceptibilities show a monotonic behaviour, the fourth order ones tend to peak around $T\sim T_c$ and the sixth order diagonal susceptibilities oscillate once from positive to negative values across the crossover temeperature. In~\cite{electricscaling} it was argued that the electric charge fluctuations have similar features as that of the baryon number fluctuations. We confirm this expectation for the $\l2+1\r$ flavor PQM in this work. In case of the fourth order fluctuations, the models do not show as pronounced a peak as lattice data, especially for $\chi^4_S$, where the slow melting of the strange condensate slows down the transition in the strange sector and produces a broad peak. For the sixth order fluctuations, the amplitude of the oscillation in the transition region is smaller compared to that of LQCD. The reasons behind these mismatch in case of fourth and sixth order fluctuations could be twofold: Firstly, currently LQCD data available for these quantities are only on a coarse lattice with a heavy pion. One has to wait for the continuum measurement with physical pion mass to make definite statements. Secondly, fluctuations of the field configurations could be playing a more significant role for these higher order susceptibilities than in case of the quadratic ones. Model computations beyond the mean field should be undertaken to verify these effects. We have also presented the model predictions for all the possible fourth order correlations for which there is no LQCD data yet. We find a sharp spike in case of the $BQ$ correlators in the transition region while for $BS$ and $QS$ correlators, there is a
sharp rise around the crossover followed by a gradual approach to the SB limit that is a characteristic of the slow melting of the strange condensate. There is also a hint of a double peak in case of ModelWB for these quantities. All the three correlators of the kind $\chi^{BQS}$ show a similar behaviour, there is a sharp rise around $T_c$ followed by a gradual melting.

We have also presented results for the ratios of susceptibilities. In the low $T$ limit, we showed that because of a nice factorisation in the expression of pressure in HRGM and PQM~\cite{chi4Btochi2B}, the ratios become insensitive to the mass spectrum and thus as long as the quantum numbers of the relevant lightest degree of freedom match, one expects PQM to agree with HRGM. Also, on taking the ratio unknown normalisation effects cancel in case of LQCD data. Thus overall, the ratios of susceptibilities are in a way better suited for comparison between lattice and model results than the susceptibilities themselves. We find that the model prediction for the ratio of the cumulants are in far better agreement with lattice data. Both PQM as well as LQCD show a monotonic behaviour for $\chi_{11}^{BQ}/\chi_{2}^{B}$ and $-\chi_{11}^{BS}/\chi_{2}^{B}$ while $-\chi_{11}^{BS}/\chi_{2}^{S}$ and $\chi_{11}^{QS}/\chi_{2}^{S}$ exhibit a sharp change in the value in the crossover region. PQM also produces a sharp peak and a plateau for the ratios $\chi_{11}^{BQ}/\chi_{2}^{Q}$ and $\chi_{11}^{QS}/\chi_{2}^{Q}$ respectively as seen in LQCD. All the fourth and sixth order normalised diagonal susceptibilities start from unity at low $T$ as also in HRGM and attain the SB limit in the high $T$ limit. In the transition region, while the ratios of fourth to second order show a peak, the normalised sixth order ones show oscillatory behaviour.

We conclude our investigation of the susceptibilities of the PQM model at zero chemical potentials and their comparison with available LQCD data. Overall it is encouraging to note the good agreement between the PQM model 
and LQCD data. The model successfully reproduces many of the features that are seen in LQCD data in the transition regime. Currently, continuum estimates for $\chi_2^B$, $\chi_2^Q$, $\chi_2^S$ and $-\chi_{11}^{BS}/\chi_2^S$ with physical pion masses are available from WB group. There is very good agreement between PQM and these quanitites except for $\chi_2^Q$ where in the low $T$ limit, the model prediction is much suppressed compared to LQCD because in the current computation at the mean field level pionic fluctuations that contribute dominantly to this quantity have not been included. In case of the correlations and higher order fluctuations, LQCD and model match qualitatively although the agreement is not as good as the earlier ones. Since fluctuations are expected to play an important role in these cases, model computations beyond the mean field might improve the agreement with LQCD. Also one has to wait for LQCD data from finer lattices to have a clear picture. We also found that the ratios of susceptibilities show better agreement with LQCD data than the susceptibilities themselves as unknown normalisation effects in LQCD data get cancelled. Also, their low $T$ values approach HRGM as dependence on the mass spectrum gets cancelled. Thus, our study of these generalised susceptibilities and their overall good agreement with lattice data provides a strong basis for the use of PQM as an effective model in the study of the QCD phase diagram. In the future, we would like to compute these generalised susceptibilities on the entire $\l T-\mu\r$ plane (similar studies in the baryon sector alone have been already performed in~\cite{sus_sign}) that will provide us valuable insights regarding the choice of the susceptibilities best suited to search for the CEP. We also plan to study the effect of including fluctuations on the susceptibilities.
\section{Acknowledgement}
S.C. would like to acknowledge discussions and collaborations with Sourendu Gupta for introduction to 
the subject itself. He would also like to thank B.~-J.~Schaefer for fruitful discussions on the subject. We would like to thank Rohini Godbole for guidance. K.A.M. acknowledges the financial 
support provided by CSIR, India.

\end{document}